\documentclass[aps,pre,twocolumn,groupedaddress,amsmath,amssymb]{revtex4-1} %[apsrev,revsymb]
\usepackage{amssymb,amsfonts,amsmath}
\usepackage{graphicx}
\usepackage{amsbsy}
\usepackage{revsymb}
\usepackage{xcolor}
\usepackage{dcolumn}
\usepackage{bm}

\def\C{{\@QC C}}
\def\@QC#1{\mathpalette{\setbox0=\hbox\bgroup$\rm}%
  {\egroup C$\egroup\rm\rlap{\kern0.4\wd0\vrule
  width 0.05\wd0 height 0.97\ht0 depth -0.01\ht0}%
  #1\bgroup}}

\newcommand{\bc}{\begin{center}}
\newcommand{\ec}{\end{center}}
\newcommand{\be}{\begin{equation}}
\newcommand{\ee}{\end{equation}}
\newcommand{\bi}{\begin{itemize}}
\newcommand{\ei}{\end{itemize}}

\newcommand{\mrtwo}{$\text{M(RT)}^2$ }

\newcommand{\psiswf}{\psi_{\text{\tiny SWF}}}
\newcommand{\psifswf}{\psi_{\text{\tiny FSWF}}}

\newcommand{\psiaswf}{\psi_{\text{\tiny ASWF}}}

\newcommand{\SD}{\text{SD}}

\newcommand{\volume}{\mathcal{V}}

\newcommand{\Aeeuu}{A_{ee}^{\uparrow \uparrow}}
\newcommand{\Aeeud}{A_{ee}^{\uparrow \downarrow}}
\newcommand{\Feeuu}{F_{ee}^{\uparrow \uparrow}}
\newcommand{\Feeud}{F_{ee}^{\uparrow \downarrow}}

\newcommand{\Aepuu}{A_{ep}^{\uparrow \uparrow}}
\newcommand{\Aepud}{A_{ep}^{\uparrow \downarrow}}
\newcommand{\Fepuu}{F_{ep}^{\uparrow \uparrow}}
\newcommand{\Fepud}{F_{ep}^{\uparrow \downarrow}}
\newcommand{\SDpw}{\text{SD}_{\text{pw}}}
\newcommand{\SDdft}{\text{SD}_{\text{DFT}}}

\newcommand{\Assuu}{A_{ss}^{\uparrow \uparrow}}
\newcommand{\Assud}{A_{ss}^{\uparrow \downarrow}}
\newcommand{\Fssuu}{F_{ss}^{\uparrow \uparrow}}
\newcommand{\Fssud}{F_{ss}^{\uparrow \downarrow}}

\newcommand{\Aspuu}{A_{sp}^{\uparrow \uparrow}}
\newcommand{\Aspud}{A_{sp}^{\uparrow \downarrow}}
\newcommand{\Fspuu}{F_{sp}^{\uparrow \uparrow}}
\newcommand{\Fspud}{F_{sp}^{\uparrow \downarrow}}

\newcommand{\psit}{\psi_{\text{T}}}
\newcommand{\Eloc}{E_{\text{loc}}}
\newcommand{\psijs}{\psi_{\text{JS}}}
\newcommand{\Jee}{J_{\text{ee}}}
\newcommand{\Jep}{J_{\text{ep}}}
\newcommand{\Jse}{J_{\text{se}}}
\newcommand{\Jsp}{J_{\text{sp}}}
\newcommand{\uyuk}{u_{\text{YUK}}}

\newcommand{\Jp}{J_{\text{p}}}
\newcommand{\Js}{J_{\text{s}}}

\usepackage[latin1]{inputenc} % Allows to use German special characters (?,?,...)

\begin{document}

%\title{Solid Molecular Hydrogen: Absence of Metallization through Bandgap Closure} 
\title{Metal-Insulator Transition of Solid Hydrogen \\ by the Antisymmetric Shadow Wave Function}

\author{Francesco Calcavecchia}
\email{{francesco.calcavecchia@gmail.com}}
\affiliation{{LPMMC, UMR 5493 of CNRS, Universit{\'e} Grenoble Alpes, 38042 Grenoble, France}}
\affiliation{{Institute of Physics, Johannes Gutenberg-University, Staudingerweg 7, D-55128 Mainz, Germany}}
\affiliation{{Graduate School of Excellence Materials Science in Mainz, Staudingerweg 9, D-55128 Mainz, Germany}}
\author{Thomas D. K\"uhne}
\email{tdkuehne@mail.uni-paderborn.de}
\affiliation{Dynamics of Condensed Matter, Department of Chemistry, University of Paderborn, Warburger Str. 100, D-33098 Paderborn, Germany}
\affiliation{Paderborn Center for Parallel Computing and Institute for Lightweight Design, Department of Chemistry, University of Paderborn, Warburger Str. 100, D-33098 Paderborn, Germany}

\date{\today}

\begin{abstract}
%{We present an improved shadow wave function approach to quantum Monte Carlo for large-scale fermionic systems. It is based on employing the antisymmetric shadow wave function in conjunction with the Gaussian determinant method to reduce the variance and an enhanced stochastic reconfiguration scheme to efficiently optimize the {trial} wave function, as well as refined twist averaged boundary conditions and periodic coordinates techniques.}
We revisit the pressure-induced metal-insulator-transition of solid hydrogen by means of variational quantum Monte Carlo simulations based on the antisymmetric shadow wave function. In order to facilitate studying the electronic structure of large-scale fermionic systems, the shadow wave function formalism is extended by a series of technical improvements, such as a revised optimization method for the employed shadow wave function and an enhanced treatment of periodic systems with long-range interactions. 
It is found that the superior accuracy of the antisymmetric shadow wave function results in a significantly increased transition pressure.
\end{abstract}

\maketitle

%\preprint{}




%\thanks{}
% \altaffiliation{}






\section{Introduction}

In 1935 Wigner and Huntington predicted that{, at very high pressure, } solid molecular hydrogen {will} dissociate {and become} {an atomic metallic solid} \cite{JCP.3.764}. Because of its relevance to astrophysics \cite{1995Sci...269.1252A}, but in particular due to 
%\dfc{the possible existence of a metallic liquid ground state \cite{2004Natur.431..669B} and high-$T_{c}$ superconductivity \cite{PhysRevLett.21.1748}}
{the possible high-$T_{c}$ superconductivity \cite{PhysRevLett.21.1748} and the existence of a metallic liquid ground state \cite{2004Natur.431..669B}}
, the importance to grasp metallic hydrogen can hardly be overstated \cite{RevModPhys.52.393, RevModPhys.66.671}. 
%In 1935 Wigner and Huntington \cite{JCP.3.764} predicted that at very high pressure solid molecular hydrogen would dissociate and form an atomic solid that is metallic. In spite of its apparent simplicity and intense research efforts the experimental realization, as well as the theoretical determination of the crystal structure has remained elusive. 
Due to the fact that it is still not possible to reach the static compression ($>$ 450 GPa) required to dissociate solid hydrogen, recently alternative routes to metallic hydrogen, though at lower pressure have been proposed \cite{RevModPhys.84.1607}. On the one hand, the negative slope of the melting line \cite{2003PNAS..100.3051S} immediately suggests the possibility of producing liquid metallic hydrogen at reduced pressure, when exposed to finite temperature \cite{PhysRevLett.100.155701, LTP.35.318, 2009JETPL..89..174E, Eremets2011}. On the other hand, due to the persistence of the molecular phase, it has been predicted that metallization through bandgap closure may be possible even in the paired state \cite{PhysRevLett.34.812, PhysRevLett.62.1150}, which would be very consequential since it facilitates potential high-$T_{c}$ superconductivity in molecular metallic hydrogen \cite{1989Natur.340..369B, PhysRevLett.78.118}. 
%However, both avenues are complicated by the fact that contrary to Phase I ($<$ 110 GPa), which is a molecular solid that is made of quantum rotors on a hcp lattice, the structures of Phase II and Phase III ($>$ 150 GPa) are still unknown \cite{PhysRevLett.61.857, PhysRevLett.63.2080}, in particular whether or not the latter is metallic \cite{1989Sci...244.1462M, PhysRevLett.76.1663, PhysRevLett.76.1667}. Since even the combined power of experimental vibrational and scattering data have not yet allowed for a unique crystal structure determination, only by means of theory a large variety of different structures have been predicted as potential candidates for Phase III, many of which were indeed metallic \cite{PhysRevLett.34.812, PhysRevLett.62.1150, PhysRevB.36.2092, PhysRevLett.66.64, PhysRevLett.67.1138, Alavi1998, PhysRevLett.83.4097, 2000Natur.403..632J, PhysRevLett.84.6070}.
However, computational studies recently demonstrated that even though the pairing structure is indeed persistent over the whole pressure range of Phase III, {it is more importantly throughout insulating} \cite{2007NatPh...3..473P, azadi2012absence, 2013NJPh...15k3005A, 2014PhRvB..90a4110S}. This is to say that metallization due to dissociation into atomic solid hydrogen may precede eventual bandgap closure.

{Thus, in} this paper, we investigate the molecular-atomic metal-insulator transition in solid hydrogen. 
%The fact that the pairing structure is persistent over the whole pressure range, and contrary to earlier theoretical predictions \cite{2000Natur.403..632J}, but in agreement with experiment \cite{2002Natur.416..613L} insulating, suggests that metallization due to dissociation may precede eventual bandgap closure. 
{Due} to the small energy differences between the various phases of high-pressure hydrogen, instead of the effective single-particle density functional theory (DFT) \cite{RevModPhys.61.689, RevModPhys.71.1253}, the more accurate Quantum Monte Carlo (QMC) method is employed here \cite{RevModPhys.73.33, LuchowQMCreview, QMCrpp2011, LesterQMCreview}. 

The remainder of the paper is organized as follows. In section~\ref{sec:VMC} we outline the variational Monte Carlo method and introduce the shadow wave function, as well as its antisymmetric variant. Section~\ref{sec:CompDet} contains the computational details, whereas in section~\ref{sec:ExtSys} we describe our implementation for extended systems. The eventual results are discussed in section~\ref{sec:results}. The last section is devoted to the conclusions.

\section{Variational Monte Carlo} \label{sec:VMC}

Variational Monte Carlo (VMC) \cite{PhysRev.138.A442}, is a QMC method that permits to approximately solve the many-body Schr\"odinger equation. The main concepts underlying VMC are the application of the Rayleigh-Ritz variational principle and importance sampled Monte Carlo (MC) to efficiently evaluate high-dimensional integrals in order to compute %{many different expectation values such as} 
the total energy \cite{Kalos:MC, Binder:MC}. 
%\begin{equation}
%  E = \frac{\int d\mathbf{R} \, \psi^{*}(\mathbf{R})H\psi(\mathbf{R})}{\int d\mathbf{R} \, \psi^{*}(\mathbf{R}) \psi(\mathbf{R})}
%\end{equation}
%VMC requires an initial ansatz that determines the quality of the results: The trial wave function.
However, in contrast to quantum-chemical electronic structure methods \cite{PopleQC}, where the computational complexity grows rapidly with the number of electrons $N$, the {formal} scaling of {VMC} is similar to that of effective single-particle theories such as Hartree-Fock (HF) or DFT \cite{HelgakerBook}. Furthermore, as many-body correlation effects are explicitly taken into account by a prescribed trial wave function (WF), VMC is throughout more accurate than typical mean-field techniques and allows to treat even strongly correlated systems. 

%Nevertheless, as the exact wave function is unknown from the outset, the trial WF ought to resemble it as closely as possible. 
Nevertheless, since the exact WF of the electronic ground state is generally unknown, it is approximated by a trial WF $\psit(R,\alpha)$, where $R \equiv \left( \mathbf{r}_1, \mathbf{r}_2, \dots, \mathbf{r}_N \right)$ are the particle coordinates. %and $\alpha \equiv (\alpha_i)_{i=1,\dots n}$ the so-called variational parameters.
The variational parameters $\alpha \equiv (\alpha_i)_{i=1,\dots n}$, which corresponds to the lowest variational energy
\begin{equation}
	E=\frac{\int dR \, \psit^*(R,\alpha) H \psit(R,\alpha) }{\int dR \, \psit^*(R,\alpha) \psit(R,\alpha) }, \label{eq:E_for_VP}
\end{equation}
represents the best possible approximation of the electronic ground state within the given trial WF, while $H$ is the system's Hamiltonian. Therefore, the accuracy of a VMC simulation depends critically on how well the particular trial WF mimics the exact ground state WF. 

For the purpose to efficiently evaluate the high-dimensional integral of Eq.~\eqref{eq:E_for_VP}, it is convenient to rewrite it as
\begin{equation}
	E=\frac{\int dR \, |\psit(R,\alpha)|^2 \frac{H \psit(R,\alpha)}{\psit(R,\alpha)} }{\int dR \, |\psit(R,\alpha)|^2  }.
\end{equation}
This facilitates to compute $E$ using the MC method by sampling $M$ points from the probability density function
\begin{equation}
	\rho(R)= \frac{|\psit(R,\alpha)|^2}{\int dR \, |\psit(R,\alpha)|^2}.
\end{equation}
Employing the \mrtwo algorithm (also known as the Metropolis algorithm) \cite{MRT2}, the variational energy can be estimated as
\begin{equation}
	E \simeq \frac{1}{M} \sum_{i=1}^M \Eloc(R_i),
\end{equation}
where
\begin{equation}
	\Eloc(R) \equiv \frac{H \psit(R,\alpha)}{\psit(R,\alpha)}
\end{equation}
is the so-called local energy. 

Even though appending a simple Jastrow correlation function to the trial WF enables to recover most of the dynamic correlation effects \cite{PhysRev.98.1479}, %Moreover, VMC is a necessary sub-step for more sophisticated methods, such as Diffusion Monte Carlo or Green's Function Monte Carlo.
we are considering the shadow wave function (SWF) of Kalos and coworkers \cite{PhysRevLett.60.1970, PhysRevB.38.4516}, as our trial WF. Its main advantage is that it allows to accurately describe 
%\dfc{all possible condensed phases (gas, liquid and solid) and even phase coexistence within the same functional form}
{localized and delocalized phases within the same functional form}
 \cite{PhysRevLett.72.2589}.
%\dfc{Hence, it is for example possible to simulate a solid without \textit{a priori} knowledge of its crystal structure, which instead emerges from the calculation itself}
{Hence, it is possible to use the same wave function for describing both insulating and metallic electronic structures}.
In addition, it even admits to compute inhomogeneous systems \cite{PhysRevB.56.5909, PhysRevB.69.024203, dandrea:fswf}.
{Finally}, the SWF has additional advantageous properties, such as for instance that many-body correlations are taken into account and that it obeys a strong similitude with the exact ground state WF \cite{KalosReatto1995}.
%Since fermions must obey Fermi-Dirac statistics to comply with the Pauli exclusion principle, an antisymmetric version of the SWF is required that changes its sign upon interchanging any two like-spin particles. 

\subsection{Shadow Wave Function}

The SWF formalism allows to systematically improve an arbitrary trial WF $\psit$ by applying the imaginary-time propagator $e^{- \tau H}$ that projects $\psit \not\perp \psi_{\text{GS}}$ onto the ground state WF $\psi_{\text{GS}}$. In order to demonstrate this, let us decompose the trial WF into
\begin{equation}
	\psit = \sum_{n=0}^{+ \infty} c_n \phi_n,
\end{equation}
where $\phi_n$ are the eigenfunctions of the Schr\"odinger equation %, i.e. $H\phi_n=E_n\phi_n$ for all $n~\in~\mathbb{N}$, with $E_n$ indicating the associated energy eigenenvalues. 
and $c_n$ the corresponding expansion coefficients. 
Employing the imaginary-time propagator onto $\psit$, we obtain
\begin{equation}
	e^{- \tau H} \psit = \sum_{n=0}^{+ \infty} c_n e^{- \tau E_n} \phi_n.
\end{equation}
The projector $e^{-\tau H}$ implies that all excited components are exponentially decaying~\footnote{If some energy eigenvalues $E_n$ are negative, the corresponding term is exponentially increasing instead of decaying. Nevertheless, it is always possible to add an appropriately chosen constant energy-shift to the Hamiltonian $H$, so that all excited components are again exponentially decaying.}, so that eventually the ground state energy $E_0$ is projected out, i.e.
\begin{equation}
	\lim_{\tau \rightarrow \infty} e^{- \tau H} \psit = \lim_{\tau \rightarrow \infty} \sum_{n=0}^{+ \infty} c_n e^{- \tau E_n} \phi_n \propto \phi_0 .
\end{equation}
From this it follows that $\psit(R)$ can be systematically improved by %(the normalization factors are inessential and therefore omitted)
\begin{subequations}
\begin{eqnarray}
	e^{-\tau H} \psit(R) & = & \langle R | e^{-\tau H} | \psit \rangle \\
	                        & = & \int dS \, \langle R | e^{-\tau H} | S \rangle \langle S | \psit \rangle,
\end{eqnarray}
\end{subequations}
where we have introduced an integral over a complete set of Dirac deltas $| S \rangle$ and omitted the inessential normalization factor. 
Assuming that $\tau \ll 1$, we now use the Trotter formula to approximate
\begin{equation}
	e^{- \tau (K + V)} \sim e^{-\frac{\tau}{2} V} e^{-\tau K}  e^{-\frac{\tau}{2} V},
\end{equation}
where $K$ and $V$ are the operators corresponding to the kinetic and potential energies, respectively \cite{Trotter}. Using the identity 
\begin{equation}
	\langle x | e^{- \tau K} | y \rangle = \frac{e^{-\frac{(x-y)^2}{4 \tau}}}{a},
\end{equation}
where $a$ is a normalization factor, the eventual expression for the improved trial WF reads as
\begin{equation}
	e^{-\tau H} \psit(R) = e^{- \frac{\tau}{2} V(R)} \int dS \, e^{- \frac{\tau}{2} V(S)} e^{- \frac{(R-S)^2}{4 \tau} } \langle S | \psit \rangle. \label{eq:first_way_to_derive_swf_final_eq}
\end{equation}

Yet, throughout our derivation we have assumed that $\tau \ll 1$, which causes that the imaginary-time propagation is rather short and the trial WF only slightly improved. In order to elongate the propagation in imaginary-time and to solve the Schr\"odinger equation exactly, the described procedure needs to be applied repeatedly, which eventually results in a formalism rather similar to the path-integral approach \cite{Feynman:PI, Kleinert:PI}. 
{However, there is no explicit importance sampling in path-integral MC methods \cite{CeperleyAlderQMC}.}
{Thus, following our original objective to find an improved and computational efficient trial WF, we rather truncate the projection after one step and refine the obtained functional form variationally. In other words, instead of approaching the limit $\tau \rightarrow 0$, we substitute $\tau$ by a variational parameter $C$ in the gaussian term. Furthermorer, we interpret the exponential $e^{-V(R)}$ as the Jastrow correlation factor $\Jp(R)$ for the protons and likewise $e^{-V(S)}$ as the corresponding two-body correlation term $\Js(S)$ for the shadows. The definition \mbox{$\langle S | \psit \rangle = \psit(S)$} entails that the original trial WF has to be evaluated on the shadow coordinates $S\equiv \left( \mathbf{s}_1, \mathbf{s}_2, \dots, \mathbf{s}_N \right)$. The latter is particularly important for the term that determines the symmetry of the SWF, which corresponds to a product of orbitals for a bosonic and a Slater-Determinant (SD) for a fermionic system, respectively \cite{PhysRev.34.1293}. As a consequence, any trial WF $\psit$ can be systematically improved by shadow formalism. The resulting SWF for a bosonic system then reads as 
\begin{equation}
	\psiswf(R) = \Jp(R) \int dS \, e^{-C \sum_{i=1}^{N} \left( \mathbf{r}_i - \mathbf{s}_i \right)^2 } \Js(S) \psit(S), \label{eq:SWF_first_way}
\end{equation}
where $\exp{\left(-C \sum_{i=1}^{N} ( \mathbf{r}_i - \mathbf{s}_i )^2\right)}=\Xi_{es}$ is the kernel that connects the electronic coordinates with the associated shadows by means of a gaussian term. 
%where $\Jp(R)$ is the Jastrow correlation factor for the protons and likewise $\Js(S)$ the corresponding two-body correlation term for the shadows. 
From the discussion above{,} it is apparent that the SWF can also be thought of as an one-step variational path-integral \cite{RevModPhys.67.279}.

\subsection{Shadow Wave Function for Fermionic Systems}
%\subsection{Antisymmetric Shadow Wave Function}

Since electrons are spin-1/2 fermions, Fermi-Dirac statistics dictates that the WF must obey the antisymmetry requirement to comply with the Pauli exclusion principle. Thus, a fermionic version of the SWF requires dealing with antisymmetric functions that are changing its sign upon interchanging any two like-spin particles, but whose nodes are inherently unknown. 
%where $\text{SD}(\mathbf{R})$ is a Slater determinant that satisfies the antisymmetry condition by changing sign upon the exchange of any two like-spin fermions \cite{PhysRev.34.1293}. 

The most natural way to devise an antisymmetrized SWF is to introduce a SD for each of the spins as a function of $S$, i.e. $\det(\phi_{\alpha}(\mathbf{s}_{\beta}^{\uparrow}))$ and $\det(\phi_{\alpha}(\mathbf{s}_{\beta}^{\downarrow}))$, where $\phi_{\alpha}$ are single-particle orbitals that are typically determined by mean-field theories, such as HF or DFT. 
%By directly applying the definition in Eq.~\eqref{eq:SWF_first_way}, we obtain the so-called Fermionic Shadow Wave Function (FSWF)
This results in the so-called Fermionic Shadow Wave Function (FSWF) 
\begin{eqnarray}
	\psifswf(R) &=& \Jee(R) \Jep(R,Q) \int dS \, e^{-C(R-S)^2} \Jse(S,R)  \nonumber \\
	            &\times& \Jsp(S,Q) \det(\phi_{\alpha}(\mathbf{s}_{\beta}^{\uparrow})) \det(\phi_{\alpha}(\mathbf{s}_{\beta}^{\downarrow})), 
\end{eqnarray}
where $\alpha$ and $\beta$ are the row and column indexes of the SDs for the spin-up and spin-down electrons, $\Jse(S,R)$ the electron-shadow and $\Jsp(S,Q)$ the shadow-proton Jastrow correlation factor \cite{KalosReatto1995, PederivaFSWF, calcavecchia:junq_paper, sign_problem}, while $Q\equiv(\bm{q}_1,\bm{q}_2,\dots,\bm{q}_M)$ are the coordinates of all $M$ protons. However, the FSWF is plagued by a sign problem \cite{calcavecchia:junq_paper, sign_problem, Calcavecchia2016}, which differs from the infamous fermion sign problem of projection QMC methods such as Green's function or diffusion MC \cite{Kalos:1974kx, Ceperley:1980vn}, but limits its applicability to relatively small systems. 

A simple \textit{ansatz} to circumvent the sign problem is the Antisymmetric Shadow Wave Function (ASWF)
\begin{eqnarray}
	\psiaswf(R) &=& \Jee(R) \Jep(R,Q) \det(\phi_{\alpha}(\mathbf{r}_{\beta}^{\uparrow})) \det(\phi_{\alpha}(\mathbf{r}_{\beta}^{\downarrow})) \nonumber \\
	            &\times& \int dS \, e^{-C(R-S)^2} \Jse(S,R) \Jsp(S,Q), 
\end{eqnarray}
where $\det(\phi_{\alpha}(\mathbf{r}_{\beta}^{\uparrow}))$ and $\det(\phi_{\alpha}(\mathbf{r}_{\beta}^{\downarrow}))$ are SDs as a function of the electronic coordinates \cite{PhysRevB.53.15129}. Even though the ASWF already includes many-body correlation effects of any order, the FSWF is superior since it accounts not only for symmetric, but, in addition, also for backflow correlation effects \cite{FeynmanBackflow, PhysRev.102.1189}. %anti-symmetrical level, whereas the ASWF improves only the symmetrical part of the WF. For further details refer to \cite{sign_problem}.

\subsection{Trial Wave Functions}

%{Beside the aforementioned SWFs we have also employed more simplistic trial WFs for comparison.}
{We now introduce the trial wave functions that we have employed in our calculations.}
In particular, the so-called Jastrow-Slater (JS) WF consists of a single SD that is multiplied by a simple Jastrow correlation factor to recover most of the dynamic correlation effects \cite{Bijl1940, Dingle1949, PhysRev.98.1479}: 
\begin{equation}
	\psijs (R) \equiv \det(\phi_{\alpha}(\mathbf{r}_{\beta}^{\uparrow})) \det(\phi_{\alpha}(\mathbf{r}_{\beta}^{\downarrow})) \, \Jee(R) \Jep(R,Q),
\end{equation}
where $\Jee$ and $\Jep$ are the Jastrow correlation factors $J=e^{-\sum_{i,j} u(r_{ij})}$ for the electron-electron and electron-proton interactions, whereas $u(r_{ij})$ is a two-body pseudopotential.

For the latter, here we have {chosen} the Yukawa-Jastrow pseudopotential for $\Jee$ and $\Jep$, respectively, which is defined as
\begin{equation}
	\uyuk(r) \equiv A \frac{1 - e^{-Fr}}{r},  
\end{equation}
where $A$ and $F$ are both variational parameters. The Yukawa-Jastrow pseudopotential is able to satisfy Kato's cusp condition from the outset, since
\begin{equation}
	\uyuk(r) \xrightarrow{\scriptscriptstyle r \to 0} AF - \frac{A F^2}{2}r + \mathcal{O}(r^2).
\end{equation}
Nevertheless, we have not utilized the cusp condition to fix one of the two parameters, but instead have determined both of them by means of the modified stochastic reconfiguration (SR) algorithm {\cite{CalcavecchiaEPL2015}}, as detailed in section~\ref{sec:CompDet}.

%{Furthermore, we have considered a JS trial WF made up of (hydrogen-like) 1s orbitals. In other words, the SD consists of single-particle WFs that has the functional form of the lowest energy solution of the Schr\"odinger equation for an isolated hydrogen atom and is parametrized by the corresponding proton position.} 
%The resulting JS-1s WF reads as
%\begin{equation}
%   {\phi_{\text{1s}}(\mathbf{r}, \mathbf{q}) = e^{-\gamma | \mathbf{r} - \mathbf{q} |},}
%\end{equation}
%{where $\gamma$ is a variational parameter.} 
%Subsequently, we have also devised another JS-type WF composed of what we call bi-atomic orbitals 
%\begin{equation}
%   {\psi_{\text{bi-atomic}}(\mathbf{r}, \mathbf{q}_1,\mathbf{q}_2) = \phi_{\text{1s}}(\mathbf{r},  \mathbf{q}_1) + \phi_{\text{1s}}(\mathbf{r}, \mathbf{q}_2)}
%\end{equation}
%{where $\mathbf{q}_1$ and $\mathbf{q}_2$ are the positions of the protons of the same $H_2$ molecule. We refer to the associated trial WF as JS-bi-atomic.}
{Moving our attention to the orbitals employed in the SD, we have considered four type of orbitals:}
\begin{enumerate}
   \item {simple plane wave (pw):}
   $$ {e^{i \mathbf{k}_i \mathbf{r}_i},} $$
   {where $\mathbf{k}_i$ are $k$-vectors in the Fermi sphere. More details about its actual implementation to include finite size effects are provided in subsection \ref{sub:tabc}.}
   \item {DFT, computed by the PWscf code of the Quantum Espresso suite of programs \cite{2009JPCM...21M5502G}. In particular, the Perdew-Burke-Ernzerhof (PBE) generalized gradient approximation to the exact exchange-correlation functional was used together with the bare Coulomb potential and an associated PW cutoff of just 8~Ry \cite{PhysRevLett.77.3865}. In order to accurately sample the first Brillouin zone, a dense {\bf k}-point mesh with at least $5^3$ special points was utilized \cite{PhysRevB.13.5188}. Again, more details are duly appropriated in subsection \ref{sub:tabc}.}
   \item {$1$s, corresponding to the lowest energy solution of the Schr\"odinger equation for an isolated hydrogen atom and is parametrized by the corresponding proton position:}
   $$  {\phi_{\text{1s}}(\mathbf{r}, \mathbf{q}) = e^{-\gamma | \mathbf{r} - \mathbf{q} |},} $$
   {where $\gamma$ is a variational parameter.}
   \item {bi-atomic, defined as}
   $$ {\psi_{\text{bi-atomic}}(\mathbf{r}, \mathbf{q}_1,\mathbf{q}_2) = \phi_{\text{1s}}(\mathbf{r},  \mathbf{q}_1) + \phi_{\text{1s}}(\mathbf{r}, \mathbf{q}_2),} $$
   {where $\mathbf{q}_1$ and $\mathbf{q}_2$ are the positions of the protons of the same $H_2$ molecule.}
\end{enumerate}

\section{Computational Details} \label{sec:CompDet}

In the following we are investigating a system comprising of $N=128$ hydrogen atoms as specified by the Hamiltonian
\begin{eqnarray}
  H &=& - \sum_{i=1}^{N} \hbar^2 \frac{\nabla^2_{i}}{2 m_e} - \sum_{I=1}^{{N}} \hbar^2 \frac{\nabla^2_{I}}{2 {N}_I} - \sum_{i,I=1}^{N} \frac{K_C}{|\mathbf{r}_{i} - \mathbf{q}_{I}|} \nonumber \\ &+& \sum_{i < j} \frac{K_C}{|\mathbf{r}_{i} - \mathbf{r}_{j}|} + \sum_{I < J} \frac{K_C}{|\mathbf{q}_{I} - \mathbf{q}_{J}|}, 
\end{eqnarray}
where $K_C=1/(4 \pi \epsilon_0)$ is the Coulomb constant and $\epsilon_0$ the electric free space permittivity. %{Given that $H$ includes the bare Coulomb potential, spin contributions are neglected. }

For the sake of simplicity, we have confined ourselves to the hcp and bcc phases as representatives for the insulating molecular and metallic atomic phases of solid hydrogen, respectively. To simulate an extended solid, 3-dimensional periodic boundary conditions (pbc) were deployed throughout, whereas the volume of the corresponding unit cell was determined by the Wigner-Seitz radius $ r_s = \sqrt[3]{3/(4\pi \rho)}$, with $\rho$ being the particle density. 

The electronic Schr\"odinger equation is approximately solved by VMC in conjunction with the various trial WFs described above using the HswfQMC code \footnote{https://github.com/francesco086/HswfQMC}. Since it is well known that conducting a QMC calculation by displacing all particles concurrently from a flat distribution entails a rather strong autocorrelation, here we have elected to use single-particle moves instead. This is to say that $\bm{r}_l^{\text{new}} = \bm{r}_l^{\text{old}}+\Delta(\eta_1,\eta_2,\eta_3)$, where $l$ is the index of the moved electron and $\Delta$ the corresponding magnitude of the displacement, while $\eta_i$ are random numbers from the interval $[-1/2,+1/2]$. Whereas {efficiently} recomputing the Jastrow correlation factor {after a single particle move} {efficiently} is relatively straightforward, this is not the case for the update of the SD. Following Ceperley et al. \cite{PhysRevB.16.3081},  
\begin{equation}
	\SD^{\text{new}} = \SD^{\text{old}} \, \sum_{j} \left( A^{-1} \right)_{j l}^ {\text{old}} \, A_{l j}^ {\text{new}},
\end{equation}
where $A$ is the matrix that generates the SD, i.e $\det(A)=\SD$. Similarly, also the inverse matrix $\left( A^{-1} \right)$ can be conveniently updated by means of 
\begin{equation}
	\left\{ \begin{array}{lcl}
	\left( A^{-1} \right)_{i l}^{\text{new}} &=& \left( A^{-1} \right)_{i l}^{\text{old}} \frac{\SD^{\text{old}}}{\SD^{\text{new}}} \\
	\left( A^{-1} \right)_{i j}^{\text{new}} &=& \left( A^{-1} \right)_{i j}^{\text{old}} - \left( A^{-1} \right)_{i l}^{\text{old}} \frac{\SD^{\text{old}}}{\SD^{\text{new}}} \\ &\times& \sum_s \left( A^{-1} \right)_{s j}^{\text{old}} A_{l s}^{\text{new}},
	\end{array} \right.
\end{equation}
with $j \neq l$. At the beginning of each VMC simulation, we set $\Delta$ so as to realize an acceptance rate of $\sim 50 \%$. Moreover, in order to reduce the autocorrelations, $3N/2$ single-particle moves were attempted between every successive evaluation of the estimators. 

%Due to the fact that the electrons of the \Htwo molecules possess antiparallel spins, it is possible to replace the SD by the orbitals themselves. To that extend we have considered two possibilities. The first one is to use simple translational invariant plane waves (pw) orbitals, by setting $\phi=1$ (since $\mathbf{k}_1$=0). In this way, only the Jastrow correlation factor accounts for all the relevant physics. Alternatively, more accurate orbitals can be computed at the DFT level. 
%{The orbitals of the SD can be either made of plane-waves (PW), or calculated at the DFT level. The latter has been computed by the PWscf code of the Quantum Espresso suite of programs \cite{2009JPCM...21M5502G}. In particular, the Perdew-Burke-Ernzerhof (PBE) generalized gradient approximation to the exact exchange-correlation functional was used together with the bare Coulomb potential and an associated PW cutoff of just 8~Ry \cite{PhysRevLett.77.3865}. In order to accurately sample the first Brillouin zone, a dense {\bf k}-point mesh with at least $5^3$ special points was utilized \cite{PhysRevB.13.5188}. }

Even though the high-dimensional integral of Eq.~\eqref{eq:E_for_VP} can be efficiently computed using the \mrtwo algorithm, it is nevertheless essential to determine the optimal variational parameters $\alpha$ that minimizes the variational energy. For that purpose we utilize the recently proposed modified SR algorithm \cite{CalcavecchiaEPL2015}, originally proposed by Sorella \cite{PhysRevB.71.241103}. Specifically, the SR method prescribes that the variational parameters are varied according to
\begin{equation}
	\delta\alpha_l = \lambda \sum_{k=1}^{n} f_k \left( s^{-1} \right)_{kl}, \label{eq:SR-direction}
\end{equation}
where %$s_{lk} = \langle O_k O_l \rangle - \langle O_l \rangle \langle O_k \rangle$, $f_k = \langle H \rangle \langle O_k \rangle - \langle O_k H \rangle$, $O_k = \frac{\partial}{\partial \alpha_k} \ln(\psit)$
\begin{equation}
	\left\{
	\begin{array}{lcl}
	s_{lk} & = & \langle O_k O_l \rangle - \langle O_l \rangle \langle O_k \rangle \\
	f_k & = & \langle H \rangle \langle O_k \rangle - \langle O_k H \rangle \\
	O_k & = & \frac{\partial}{\partial \alpha_k} \ln(\psit)
	\end{array}
	\right\}
\end{equation}
and $\langle \cdot \rangle \equiv \langle \psit | \cdot | \psit \rangle$. Once the gradient in the variational parameters space, which minimizes the variational energy has been computed, the step length $\lambda$ along this direction needs to be identified. Since determining the new direction $\delta \alpha$ is computational approximately equally expensive than calculating the the variational energy, it is convenient to start with a rather small value for $\lambda$ and continuously adjusting it on the fly during the optimization. %To that extent, $\lambda$ is decreased whenever the search direction changes, which means that we are ``moving too fast'', and increased otherwise. 

%\section{Antisymmetric Shadow Wave Function for extended systems}
\section{Variational Monte Carlo for extended systems}
\label{sec:ExtSys}

%\dfc{Nevertheless, employing the ASWF to study extended systems}
{When dealing with extended systems}, special care is required to accurately consider pbc and single-electron finite size effects.

\subsection{Periodic Coordinates}

If computed in its straightforward fashion, the Yukawa-Jastrow, as any other slowly decaying Jastrow correlation factor, leads to a spurious bias in the kinetic energy. Therefore, all contributions that are originating from the periodic images of the unit cell must be taken explicitly into account in order to avoid discontinuities in the derivatives of the WF when the particle distances switch from one closest image to the other. Needless to say that this approach is computationally {relatively time} consuming and a more economic strategy {very} desirable. 

However, before presenting our solution to this effect, let us start by introducing a particular useful test to verify if all correlations %\dfc{regarding the kinetic energy within periodic boundary conditions are correctly taken into account}
{are correctly taken into account}.
%To be more precise, the fulfillment of such a test is condition necessary but not sufficient.
To that extent, the expression for the kinetic energy (for simplicity we consider the kinetic contribution of only one particle $j$) is integrated by parts 
\begin{eqnarray}
	&& \int_{\Omega} dR \, \psi^*(R) \nabla_j^2 \psi(R) = \sum_{x_{\alpha}} \int_{\Omega} dR \, \psi^*(R) \left( \frac{\partial^2}{\partial x_{\alpha}^2} \right) \psi(R) \nonumber \\ 
	&&= \sum_{x_{\alpha}} \int_{\Omega} dR^{\bar{x}_{\alpha}} \, \int_{-{L_{x_{\alpha}}}/2}^{+{L_{x_{\alpha}}}/2} dx_{\alpha} \, \psi^*(R) \frac{\partial^2}{\partial x_{\alpha}^2} \psi(R) \nonumber \\
	&&= \sum_{x_{\alpha}} \left\{ \int_{\Omega} dR^{\bar{x}_{\alpha}} \, \left[ \psi^*(R) \frac{\partial}{\partial x_{\alpha}} \psi(R) \right]_{-{L_{x_{\alpha}}}/2}^{{+L_{x_{\alpha}}}/2} \right. \nonumber \\
	&&- \left. \int_{\Omega} dR \, \left( \frac{\partial}{\partial x_{\alpha}} \psi^*(R) \right) \left(\frac{\partial}{\partial x_{\alpha}} \psi(R) \right) \right\}, 
\end{eqnarray}
where $x_{\alpha}= (x_j,y_j,z_j)$, $dR^{\bar{x}_{\alpha}}$ is the same as $dR$ but excluding the infinitesimal element $dx_{\alpha}$, $\Omega$ represents the domain of integration, i.e. the simulation cell, while $L_{x_{\alpha}}$ is the length of the edge of $\Omega$ along the $x_{\alpha}$ axis. However, at the presence of periodic boundary conditions, the WF $\psi(R)$ and also its derivatives are required to be periodic, meaning that they are invariant with respect to particle translations $\mathbf{v}=\left( n_x L_x, n_y L_y, n_z L_z \right)$, where  $n_x$, $n_y$ and $n_z$ are all integers. From this follows that the term
\begin{equation}
	\left[ \psi^*(R) \frac{\partial}{\partial x_{\alpha}} \psi(R) \right]_{-{L_{x_{\alpha}}}/2}^{+{L_{x_{\alpha}}}/2} \label{eq:JF_vanisinhg_term}
\end{equation}
vanishes, which leads to a modified Jackson-Feenberg (JF) kinetic energy expression \footnote{The original Jackson-Feenberg expression reads as: \unexpanded{
\begin{equation*}
  \frac{\hbar^2}{2} \sum_{j=1}^{N} {\frac{1}{2 m_j} \int_{\Omega} dR \, \left( \nabla_j \psi^*(R) \nabla_j \psi(R) - \psi^*(R) \nabla_j^2 \psi(R) \right)}
\end{equation*}}}
\begin{equation}
	E_{\text{JF}} = \hbar^2 \sum_{j=1}^N \frac{1}{2 m_j} \int_{\Omega} dR \, \nabla_j \psi^*(R) \cdot \nabla_j \psi(R).
\end{equation}
As a consequence, the equivalence of the Pandharipande-Bethe~(PB)
\begin{equation}
E_{\text{PB}} = -\hbar^2 \sum_{i=1}^N \frac{1}{2 m_j} \int_{\Omega} dR \, \psi^*(R) \nabla_j^2 \psi(R) 
\end{equation}
and Jackson-Feenberg~(JF) expressions for the kinetic energy is a necessary, but not sufficient condition 
{for the required periodic properties of the WF}. 
%However, the equivalence of the JF and PB kinetic energies is only a necessary, but not sufficient condition for the required properties of the wave function.
Thus, in all of our calculations we have computed both expressions and explicitly verified that both are indeed identical, within the corresponding statistical uncertainties.
%In the following we refer to this comparison as the \emph{JF test}.

However, at the presence of additional correlation terms, such as the Jastrow, Eq.~\ref{eq:JF_vanisinhg_term} must be correctly interpreted since the inter-particle distances are computed using the closest periodic image. In fact, even though the particle coordinate $\mathbf{r}_j$ is confined to the unit cell of volume $\volume=L_x L_y L_z$, the distance $\mathbf{r}_{ij}$ between the particles $i$ (assumed as fixed) and $j$ does not range within $\sqrt{\left( \mathbf{r}_j - \mathbf{r}_i \right)^2 }$, 
%\begin{equation}
%	\sqrt{\left( \mathbf{r}_j - \mathbf{r}_i \right)^2 } \qquad \text{with } \mathbf{r}_j \text{ spanning } \volume
%\end{equation}
but always within the box of volume $\volume$ centered on particle $i$. This concept is illustrated in Fig.~\ref{fig:closest_image}.
%Since we are integrating in a periodic box, it must hold the equivalence
%\begin{equation}
%	\int_{-L_x/2}^{L_x/2} dx f(x) = \int_0^{L} dx f(x) = \int_{0+a}^{L_x+a} dx f(x) \, ,
%\end{equation}
%with $a \in \mathbb{R}$ and $f:\mathbb{R} \rightarrow \mathbb{R}$ any function.
Therefore, it is possible and convenient to fix the origin at the position of the $i$-th particle that is considered. As a consequence, in the following we will set $\mathbf{r}_i=0$, so that $\mathbf{r} = \left( x,y,z \right)  \equiv \mathbf{r}_{ij}=\mathbf{r}_j$ and $r=|\mathbf{r}|=\sqrt{x^2 + y^2 + z^2}$.
\begin{figure}%[htbp]
	\centering
		\includegraphics[width=1.0\columnwidth]{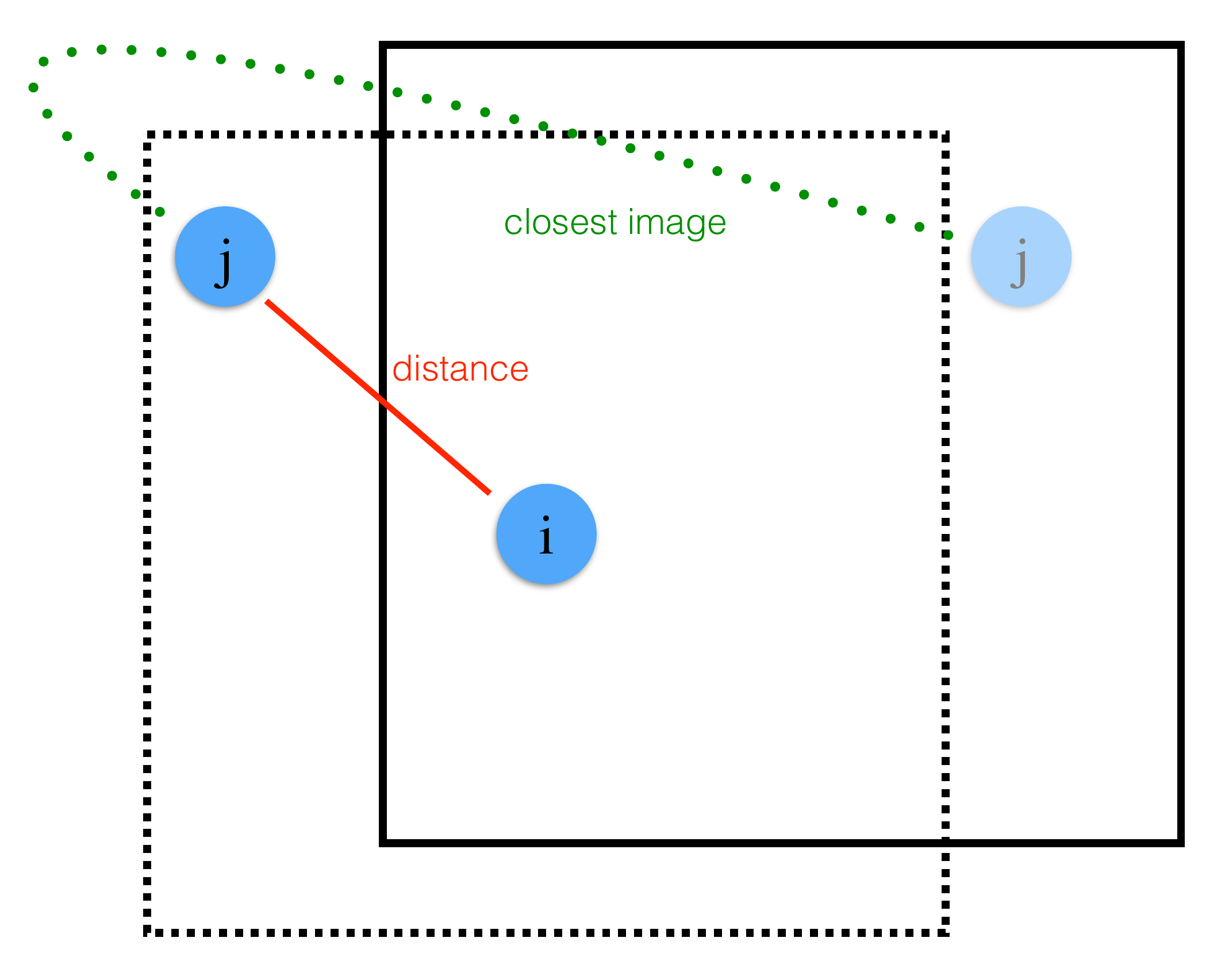}
	\caption{Integration of $\mathbf{r}_j$ within a box with periodic boundary conditions, from the point of view of particle $i$. The continuous black line represents the simulation box, whereas the dotted black line denotes the effective volume of integration for the distance between the particles $i$ and $j$.}
	\label{fig:closest_image}
\end{figure}

Let us demonstrate the JF test by showing that the Yukawa-Jastrow violates it. For that purpose we consider the simple case of only two interacting particles, i.e. 
\begin{equation}
	J(r) = e^{- \frac{A \left( 1-e^{-Fr} \right)}{r} }.
\end{equation}
Its first derivative along the $x$ axis reads as 
\begin{eqnarray}
	\frac{\partial J(r)}{\partial x} &=& \frac{\partial e^{- \frac{A \left( 1-e^{-Fr} \right)}{r} }}{\partial x} \nonumber \\
    & = & \frac{\partial e^{- \frac{A \left( 1-e^{-Fr} \right)}{r} }}{\partial r} \frac{\partial r}{\partial x} \nonumber \\
	& = & e^{- \frac{A \left( 1-e^{-Fr} \right)}{r} } \left( A \frac{1-e^{-Fr}}{r^2} - AF \frac{e^{-Fr}}{r}\right) \, \frac{x}{r} \nonumber \\
	& = & \frac{\partial J(r)}{\partial r} \frac{x}{r}.
\end{eqnarray}
It is then apparent that the first derivative is not continuous at $x= \pm L/2$, which is the border between its two closest periodic images.
As a consequence, 
\begin{equation}
	\lim_{\varepsilon \rightarrow 0} \left[ J(r) \frac{\partial}{\partial x} J(r) \right]_{-({L_x}/2) + \varepsilon}^{+({L_x}/2) - \varepsilon} = J(r) \frac{\partial J(r)}{\partial r} \frac{L_x}{r} \neq 0. \label{eq:limite_yukawa}
\end{equation}
{Therefore}, the JF and PB kinetic energies differ since the term in Eq.~\ref{eq:JF_vanisinhg_term} does not vanish. Yet, if $\frac{1}{r} \frac{\partial J}{\partial r}$ is small enough at $x=\pm \frac{L}{2}$, the difference is negligible.

An even deeper understanding can be obtained by means of the distribution theory. In fact, %one can notice that
\begin{eqnarray}
	\frac{\partial^2}{\partial x^2} J(r) &=& \frac{x}{r} \left( \frac{\partial^2 J(r)}{\partial r^2} \frac{x}{r} + \frac{\partial J(r)}{\partial r} \frac{1}{r} - \frac{\partial J(r)}{\partial r} \frac{1}{r^2} \right) \nonumber \\ 
	&-& \left( \frac{\partial J(r)}{\partial x} \frac{x}{r} \right) 2 \, \delta\left(x-\frac{L_x}{2}\right), 
\end{eqnarray}
since
\begin{equation}
	\int_{L_x/2 - \varepsilon}^{L_x/2 + \varepsilon} \frac{\partial}{\partial x} \left( \frac{\partial J(r)}{\partial x} \right) = - \frac{\partial J(r)}{\partial x} \frac{L_x}{r}.  
\end{equation}
In other words, the discontinuity in the first derivative entails a Dirac delta in the second derivative. Obviously, this artifact must be circumvented in order to avoid a bias in the computation of the kinetic energy.

Since a discontinuity in the first derivative not only affects the validity of the JF expression but also the PB one, the kinetic energy contribution provided by the Yukawa-Jastrow is biased. Mathematically, the problem can be eliminated by enforcing a smooth change between the closest periodic images. Physically, all of this originates from the fact that the simulation box is not large enough to {``}contain{''} {all} correlations between the particles.

A straightforward solution to remedy the latter is inspired by the Ewald summation technique~\cite{Natoli1995171}. More specifically, the Jastrow is decomposed into a quickly and a slowly decaying part, which are computed separately in real and reciprocal $k$-space, respectively. However, this method requires a summation over the whole momentum space, which is computationally rather demanding.

An alternative approach, which is not only more elegant and simpler, but at the same time also more efficient, is due to Attaccalite and Sorella and results from exploiting {Periodic Coordinates}~(PC)~\cite{thesis_attaccalite}. %We repute this solution to be the best option for its elegance, effectiveness and simplicity. 
As the name suggests, the only modification required is to substitute the original coordinates by 
\begin{subequations}
\begin{eqnarray}
	x^{\prime} &=& \frac{L}{\pi} \sin\left(\frac{\pi x}{L} \right), \\
	y^{\prime} &=& \frac{L}{\pi} \sin\left(\frac{\pi y}{L} \right), \\
	z^{\prime} &=& \frac{L}{\pi} \sin\left(\frac{\pi z}{L} \right),
\end{eqnarray}
\end{subequations}
and hence evaluate the distances via 
\begin{equation}
	r^{\prime} = \frac{L}{\pi} \sqrt{ \sin^2\left(\frac{\pi x}{L}\right) + \sin^2\left(\frac{\pi y}{L}\right) + \sin^2\left(\frac{\pi z}{L}\right) }.
\end{equation}
The employment of Periodic Coordinates enforces the correct periodicity of the WF. For example, the first derivative
\begin{eqnarray}
	\frac{\partial J(r^{\prime})}{\partial x} &=& \frac{\partial J(r^{\prime})}{\partial r^{\prime}} \frac{\partial r^{\prime}}{\partial x^{\prime}} \frac{\partial x^{\prime}}{\partial x} \nonumber \\
	    &=& \frac{\partial J(r^{\prime})}{\partial r^{\prime}} \frac{x^{\prime}}{r^{\prime}} \cos\left(\frac{\pi x}{L} \right),
\end{eqnarray}
is continuous in $x= \pm L/2$, i.e. on the borders of the simulation box. The same also holds for all higher order derivatives. The consequential modifications of the Yukawa Jastrow are illustrated in Fig.~\ref{fig:PC}. 
\begin{figure}%[htbp]
	\centering
		\includegraphics[width=1.0\columnwidth]{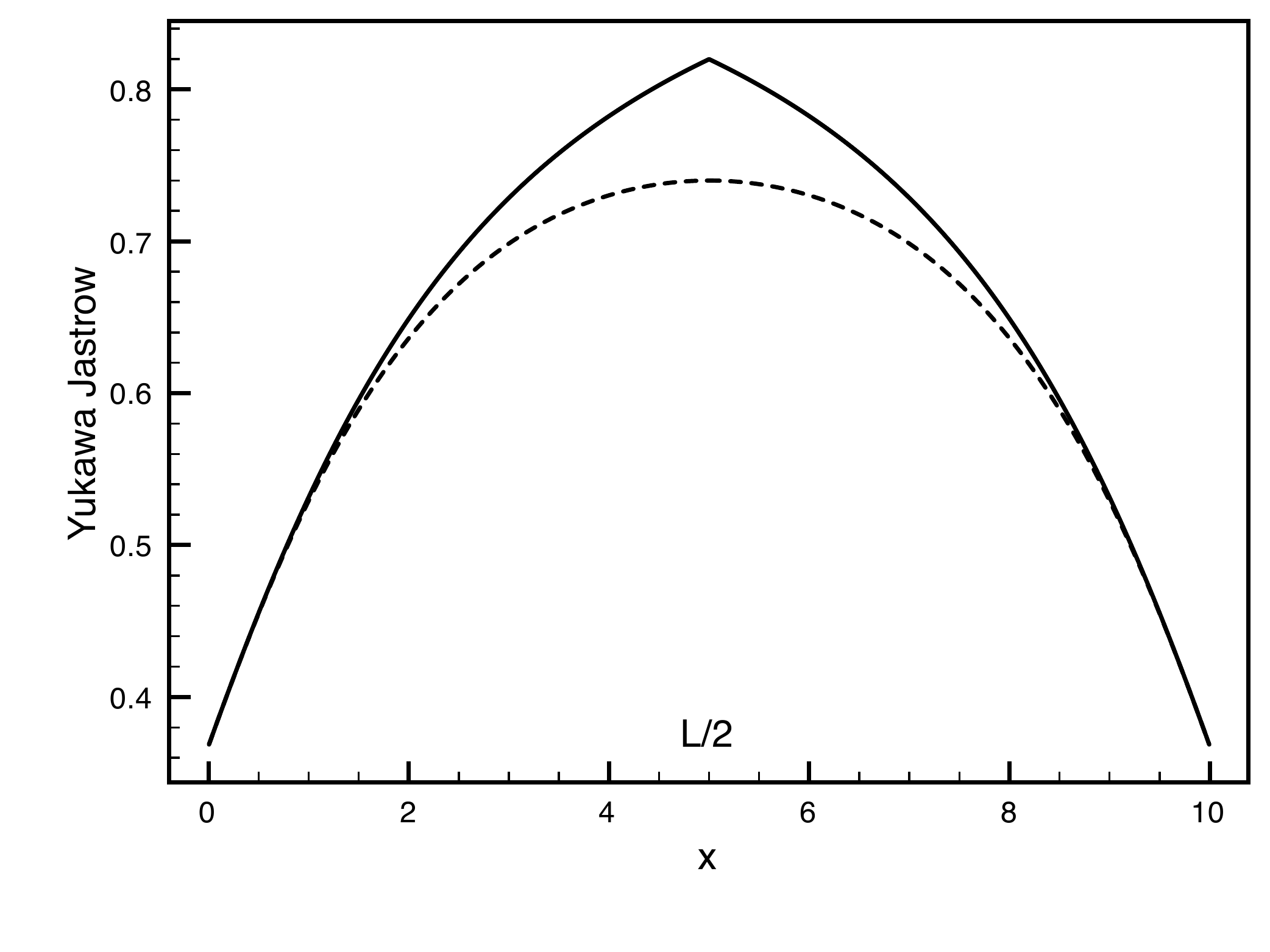}
	\caption{Comparison between the Yukawa Jastrow correlation factor for two interacting particles $\exp\left( - \frac{A \left(1-\exp(-Fr) \right)}{r} \right)$ (black line) and its modified version as obtained by employing Periodic Coordinates (dashed line). We have set $y=z=0$, $L=10$ and $A=F=1$, respectively.}
	\label{fig:PC}
\end{figure}

To demonstrate the effectiveness of PC, we have calculated the kinetic energy using the JS-pw trial WF for two different systems, each consisting of $16$ hydrogen atoms. The results of the atomic bcc (atm-bcc) and the molecular hcp (mol-hcp) phases of solid hydrogen including the corresponding Wigner-Seitz radii are shown in Table~\ref{tab:PC}. 
\begin{table}
\caption{\label{tab:PC} Kinetic energies (in Ry) for the atm-bcc and mol-hcp phases of solid hydrogen as obtained with and without PC.}
\begin{ruledtabular}
\begin{tabular}{l|c|c}
& without PC & with PC \\ \hline
\begin{tabular}{l} atm-bcc \\ $r_s=1.31$ \end{tabular} & \begin{tabular}{l} $E_{\text{kin}} = 5.5766(5)$ \\
		$E_{\text{JF}} = 2.9841(32)$ \end{tabular} & \begin{tabular}{l} $E_{\text{kin}} = 1.5480(6)$ \\
		$E_{\text{JF}} = 1.5454(16)$ \end{tabular} \\ \hline
\begin{tabular}{l} mol-hcp \\ $r_s=2.61$ \end{tabular} & \begin{tabular}{c} $E_{\text{kin}} = 2.2428(6)$ \\
		$E_{\text{JF}} = 2.1252(10)$ \end{tabular} & \begin{tabular}{c} $E_{\text{kin}} = 1.0307(11)$ \\
		$E_{\text{JF}} = 1.0290(5)$ \end{tabular}
\end{tabular}
\end{ruledtabular}
\end{table}
%We obtained 
%\begin{equation}
%	\text{[without PC]} \qquad \begin{array}{rcl}
%		E_{\text{kin}} &=& 5.5766(5) \\
%		E_{\text{JF}}  &=& 2.9841(32)
%	\end{array}
%\end{equation}
%without using the PC, whereas, by applying them, we attained 
%\begin{equation}
%	\text{[with PC]} \qquad \begin{array}{rcl}
%		E_{\text{kin}} &=& 1.5480(6) \\
%		E_{\text{JF}}  &=& 1.5454(16)
%	\end{array}
%\end{equation}
%The success is confirmed for the mol-hcp lattice at $r_s=2.61$:
%\begin{equation}
%	\text{[without PC]} \qquad \begin{array}{rcl}
%		E_{\text{kin}} &=& 2.2428(6) \\
%		E_{\text{JF}}  &=& 2.1252(10)
%	\end{array}
%\end{equation}
%\begin{equation}
%	\text{[with PC]} \qquad \begin{array}{rcl}
%		E_{\text{kin}} &=& 1.0307(11) \\
%		E_{\text{JF}}  &=& 1.0290(5)
%	\end{array}
%\end{equation}
As can be extracted by comparing $E_{\text{kin}}$ with $E_{\text{JF}}$, the aforementioned spurious bias can be completely eliminated by the use of PC with a only negligible additional computational cost.
%We point out that in all cases we used improper variational parameters, therefore their values are meaningless from a physical point of view.
Nevertheless, we find it important to remark that employing PC leads to a somewhat modified Yukawa Jastrow, which may slightly violate the electron-electron and electron-proton cusp conditions \cite{Kato}.
%This problem does not occur by implementing an Ewald decomposition rather than the PC coordinates.
However, the accuracy of employed Jastrow in conjunction with PC can be easily checked by means of the variational principle. We have explicitly verified that in practice the latter bias is generally tiny. 

\subsection{SWF Kernel Truncation}

If the variational parameter $C$ of the SWF kernel is small, the simulation box is typically not large enough to constrain each particle to its associated shadows $\mathbf{s}_1$ and $\mathbf{s}_2$ within the limit $L/2$. As before, this entails a bias in the kinetic energy, as can be seen by the difference between $E_{\text{kin}} = 1.624(4)~\text{Ry}$ and $E_{\text{JF}}  = 1.478(5)~\text{Ry}$, respectively
%comparing the Pandharipande-Bethe kinetic energy with the Jackson-Feenberg expression 
\footnote{The following estimated kinetic energies have been computed for $16$ hydrogen atoms in the metallic atm-bcc phase at $r_s=1.31$ using the ASWF-pw trial WF. The employed variational parameters are: $\Aeeuu=0.423$, $\Aeeud=0.829$, $\Feeuu=2.568$, $\Feeud=1.834$, $\Aepuu=-74.930$, $\Aepud=-68.191$, $\Fepuu=0.231$, $\Fepud=0.242$, $C=0.542$, $\Assuu=2.400$, $\Assud=2.112$, $\Fssuu=5.508$, $\Fssud=19.039$, $\Aspuu=2.400$, $\Aspud=2.112$, $\Fspuu=5.508$ and $\Fspud=19.039$, respectively\label{footnote:variational_parameters_kernel_truncation}}. 
{In order to eliminate this shortcoming, the kernel must be modified so that it vanishes for $|\mathbf{r}-\mathbf{s}| \rightarrow L/2$.} An appropriate choice for the modified kernel reads as 
\begin{equation}
	\Xi_{es}(R,S) = \prod_{i=1}^N e^{- \varsigma(|\mathbf{r}_i - \mathbf{s}_i |)},
\end{equation}
where 
\begin{equation}
	\varsigma(x) = \left\{ \begin{array}{lcc}
								C x^2 & & \text{if $x \leq \frac{L}{6}$} \\
								\alpha_0 + \frac{\alpha_1}{x-L/2}  & & \text{if $\frac{L}{6} < x \leq \frac{L}{2}-\varepsilon$} \\
								\beta_0 + \beta_1 x^n  & & \text{if $x > \frac{L}{2} - \varepsilon$}
							\end{array}
					\right. , \label{eq:kernel_modification_final}
\end{equation}
with
\begin{equation}
	\left\{ \begin{array}{l}
		\varepsilon = \frac{L}{n+1}   \\
		\beta1 = - \frac{2 \alpha_1}{\varepsilon^3 n \left( n-1 \right) \left( \frac{L}{2} - \varepsilon \right)^{n-2}  }   \\
		\beta_0 = \alpha_0 - \frac{\alpha_1}{\varepsilon} - \beta_1 \left( l - \varepsilon \right)^n  
	\end{array} \right.
\end{equation}
and $n \geq 2$.
Our simulations have suggested that a suitable choice is $n=12$. The modification introduced by Eq.~\eqref{eq:kernel_modification_final} are illustrated in Fig.~\ref{fig:kernel_exponent}. 
\begin{figure}%[htbp]
	\centering
		\includegraphics[width=1.0\columnwidth]{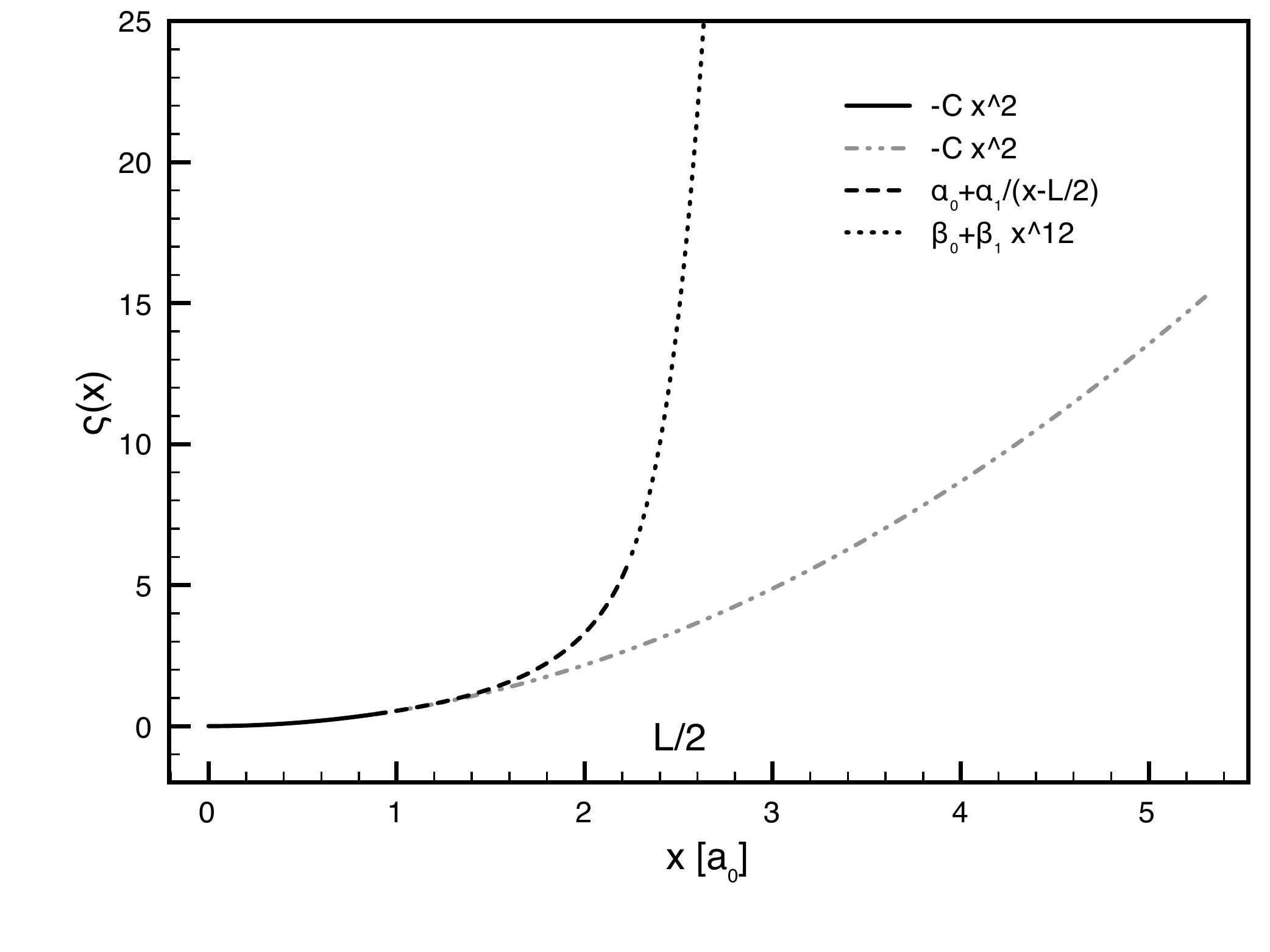}
	\caption{Illustration of the SWF kernel truncation method prescribed by Eq.~\eqref{eq:kernel_modification_final} with $C=0.542$, for 16 hydrogen atoms in the metallic atm-bcc phase at $r_s=1.31$ ($L/2 \simeq 2.66~\text{a}_0$).}
	\label{fig:kernel_exponent}
\end{figure}
The corresponding kinetic energies are $E_{\text{kin}} = 1.538(6)~\text{Ry}$ and $E_{\text{JF}} = 1.533(6)~\text{Ry}$, which demonstrates that the proposed SWF kernel truncation method completely alleviates the aforementioned limitation. 

\subsection{Twist Averaged Boundary Conditions}
\label{sub:tabc}

As already alluded to previously, the application of pbc do not automatically result in an accurate description of an infinite system. In fact, identical simulations but using distinct values for $N$ may entail rather different results. As a consequence, these effects are generally referred to as {finite-size effects}, which can be minimized %We now present a method that permits its reduction, called {Twist Averaged Boundary Conditions} (TABC)~\cite{PhysRevE.64.016702}.
by the usage of so-called {Twist Averaged Boundary Conditions} (TABC)~\cite{PhysRevE.64.016702}. %have been devised in the context of metallic systems, wherein the Slater determinant of plane waves provides a good description of the electronic structure.
The origin of these {finite-size effects} are that the embedded $\mathbf{k}$-vectors do not well represent an infinite system, since in general a discrete grid of points cannot reproduce the whole Fermi sphere (see Fig.~\ref{fig:k_fermi}). 
\begin{figure}%[htbp]
	\centering
		\includegraphics[width=1.0\columnwidth]{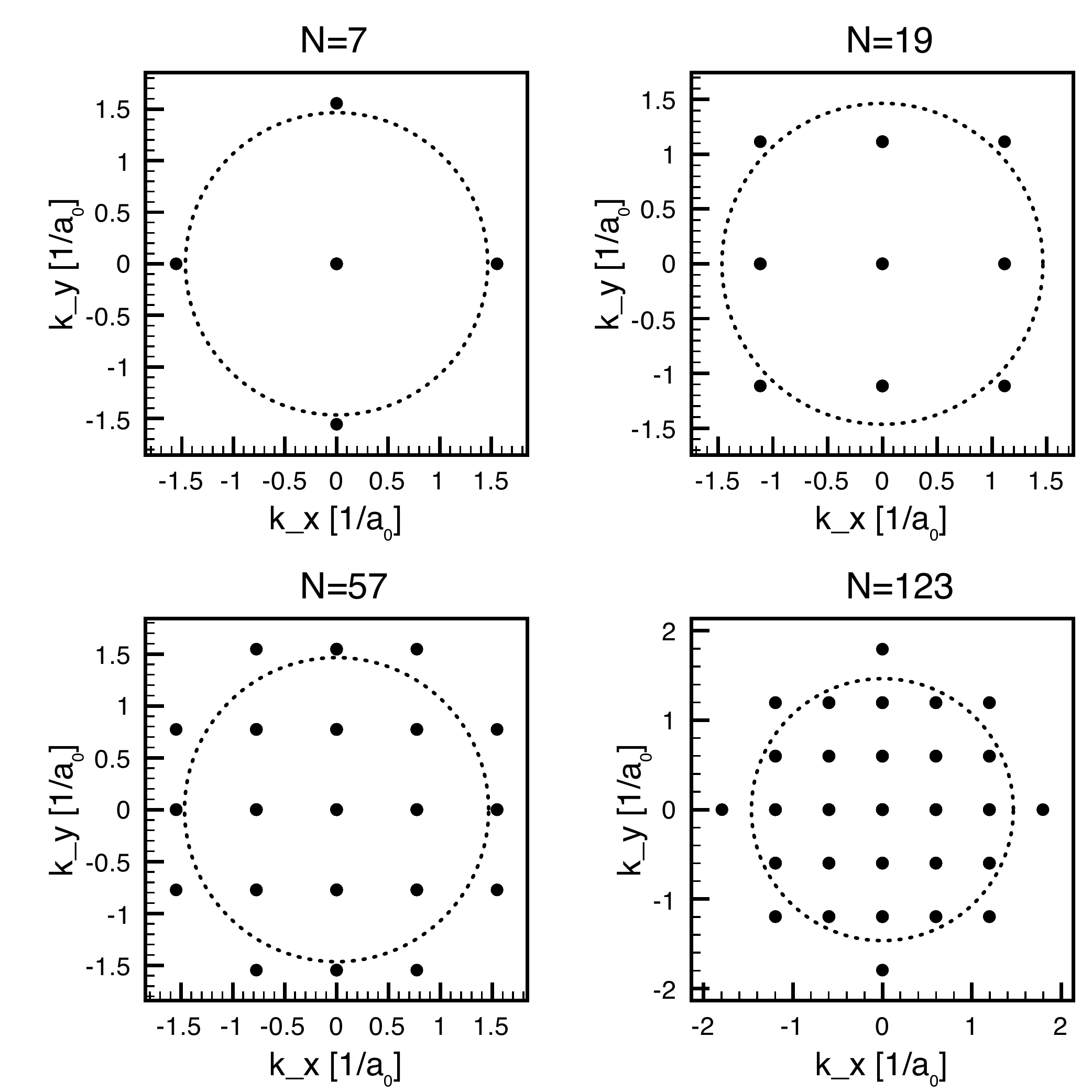}
	\caption{Two-dimensional cross section of the momenta $\mathbf{k}$ of a $N \times N$ $\SDpw$ matrix, for several values of $N$. The dotted circles delineate the Fermi sphere.}
	\label{fig:k_fermi}
\end{figure}
The TABC method, which allows to bypass this limitation by means of an integration over the Fermi sphere, prescribes a recurrent random shift
\begin{equation}
	\mathbf{v_{\text{twist}}} = \frac{2 \pi}{L} \left( \eta_1, \eta_2, \eta_3 \right)  \label{eq:v_twist}
\end{equation}
of the $\mathbf{k}$-grid, where $\eta_i $ are random numbers sampled in the range $\left[ -\frac{1}{2} , \frac{1}{2} \right] $. Within the context of TABC, the translation vector $\mathbf{v_{\text{twist}}}$ is referred to as {twist}. The corresponding integral over $\mathbf{k}$-space is computed by means of MC. As can be seen in Fig.~\ref{fig:TABC_PW}, the application of TABC results in an accelerated convergence to the thermodynamic limit. 
\begin{figure}%[htbp]
	\centering
		\includegraphics[width=1.0\columnwidth]{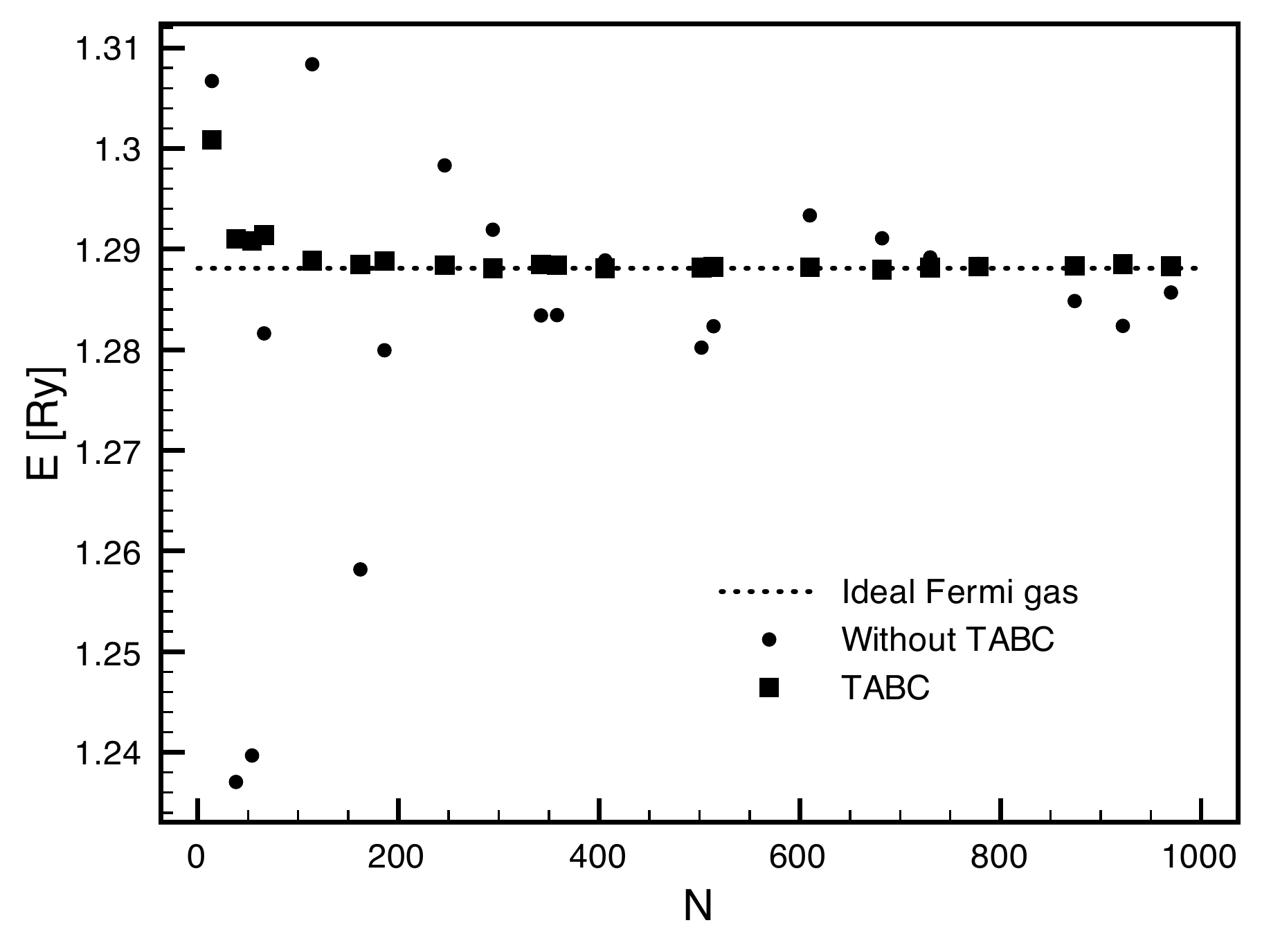}
	\caption{Electronic kinetic energy of solid hydrogen at $r_s=1.31$ as computed using $\SDpw$, where the number of particles $M$ are magic numbers~\ref{footnote:magic_numbers1}.}
	\label{fig:TABC_PW}
\end{figure}
Beyond solely reducing finite-size effects, employing TABC also permits calculations, where the number of particles $M$ are distinct from magic numbers\footnote{Magic numbers are those that close the Fermi momenta shell in a simple cubic box. For a three dimensional system these are $1,7,19,27,33,57,81,93,123,147,171,179,203,251 \dots$. \label{footnote:magic_numbers1}}, without spurious drift and anisotropy effects.

The eventual algorithm for a VMC simulation of a 3D unpolarized system employing TABC with $N_{\text{twist}}$ twists reads as follows:
\begin{enumerate}
	\item Determine the smallest magic number $n$ that is larger than $N/2$; %Examples: if $N=54$, then $n=33$, if $N=128$ then $n=93$;
	\item Find the first $n$ Fermi $\mathbf{k}$-vectors, yielding $\Gamma_0 = \left\{ \mathbf{k}_1, \mathbf{k}_2, \dots, \mathbf{k}_n \right\}$;
	\item Generate $\mathbf{v_{\text{twist}}}$ as described in Eq.~\eqref{eq:v_twist}; \label{pt:terzo_tabc1}
	\item $K = \Gamma_0 + \left\{ \mathbf{v_{\text{twist}}} \right\}_n $;
	\item Sort the $\mathbf{k}$-vectors in $K$ by increasing magnitude, and then use the first $N/2$ $\mathbf{k}$-vectors to build up $\SDpw$;
	\item Perform $M_{\text{relax}}$ relaxation steps;
	\item Sample $M/N_{\text{twist}}$ points and accumulate the estimators of the observables of interest (normally the kinetic and potential energies); \label{pt:settimo_tabc1}
	\item Repeat the points~\ref{pt:terzo_tabc1}-\ref{pt:settimo_tabc1} $N_{\text{twist}}$ times.
\end{enumerate}
The $M_{\text{relax}}$ relaxation steps of point 6 are essential to prevent the emergence of a bias in the calculation. Even though it is possible to circumvent this step by submitting the twist to the acceptance/refuse process of the \mrtwo algorithm, we have not exploited this possibility, since the number of relaxation steps is small and its computational cost negligible. 

However, as a consequence of the twist, a momentum in the external shell, which initially was not included in $\SDpw$, may indeed have a lower magnitude than the employed ones. This is to say that such a momentum actually replaces the one with the actual highest magnitude. Therefore, in step 1 of the just outlined algorithm, more $\mathbf{k}$-vectors than strictly necessary to generate $\SDpw$ are considered and eventually selected as described in point 5. The corresponding kinetic energies generated by this method are reported in Fig.~\ref{fig:effetto_twist}.
\begin{figure}%[htbp]
	\centering
		\includegraphics[width=1.0\columnwidth]{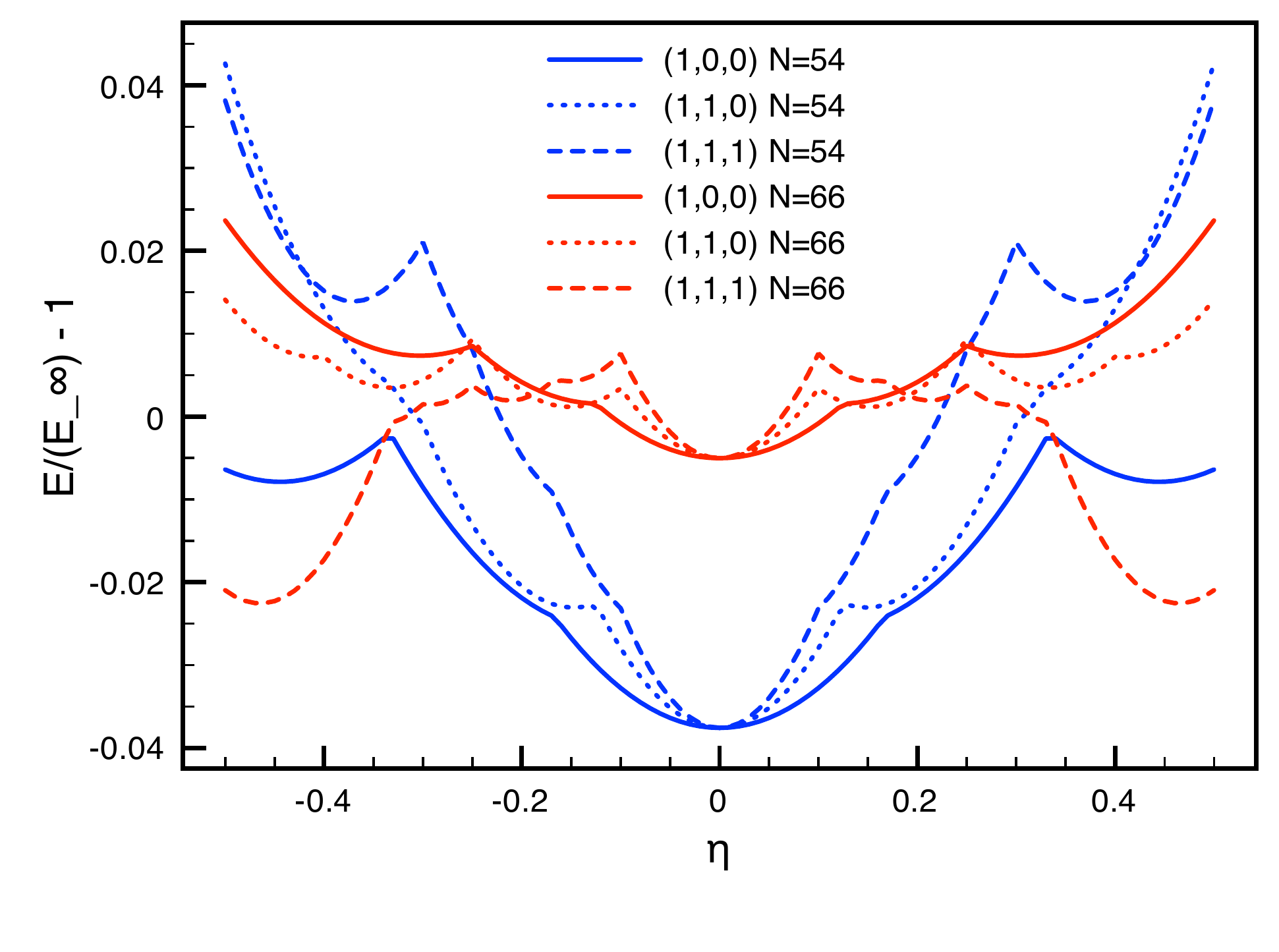}
	\caption{Kinetic energy of solid hydrogen with $54$ and $66$ particles for the twists $\mathbf{v_{\text{twist}}}=\frac{2 \pi}{L} \eta (1,0,0)$, $\mathbf{v_{\text{twist}}}=\frac{2 \pi}{L} \eta (1,1,0)$ and $\mathbf{v_{\text{twist}}}=\frac{2 \pi}{L} \eta (1,1,1)$, respectively.}
	\label{fig:effetto_twist}
\end{figure}

In the following we present our extension of the TABC approach to $\SDdft$. In fact, the DFT method itself also suffers from finite-size effects, which requires to sum over contributions from different $K$-points in the first Brillouin zone. %In particular, Quantum-Espresso allows for considering a grid of $K$ points to average over.
The simplest grid consists of just one point, denoted as $\Gamma_0$, which corresponds to the Fermi gas momenta. In order to reduce finite-size effects within DFT, it is essential to consider multiple $K$-points to yield a more accurate averaged estimate of the aforementioned integral, similarly to TABC technique. Typically, the $K$-point grids are generated using the Monkhorst and Pack construction scheme \cite{PhysRevB.13.5188}. %This can be accomplished within Quantum-Espresso by selecting a grid, for example $5 \times 5 \times 5$, and letting the algorithm find the optimal $K$-points to average over, according to the Monkhorst-Pack rule~\cite{PhysRevB.13.5188}. 
Due to the fact that each $K$-point has an associated weight, instead of summing over all weighted configurations, we propose here to adopt the TABC approach with a probability proportional to their weight. In other words, we average over all $K$-points, while making the most of importance sampling.

The implementation of the modified TABC method for $\SDdft$ can be summarized by the following instructions:
\begin{enumerate}
	\item Conduct a DFT plane-wave calculation with an energy cutoff $E_{ctf}$ in order to obtain $n_{\text{K}}$ solutions, one for each $K$-points $K_i$ and its associated weight $w_i$;
	\item Sample each $K$-point $K_j$ with probability $$P_j = \frac{w_j}{\sum_{l=1}^{n_{\text{K}}} w_l }$$ and employ its associated solutions in the $\SDdft$;\label{pt:secondo_tabc2}
	\item Perform $M_{\text{relax}}$ relaxation steps;
	\item Sample $M/N_{\text{twist}}$ points and accumulate the estimators;\label{pt:quarto_tabc2}
	\item Repeat the points~\ref{pt:secondo_tabc2}-\ref{pt:quarto_tabc2} $N_{\text{twist}}$ times.
\end{enumerate}

%In Fig.~\ref{fig:TABC-E_ctf} we report the results obtainable employing the TABC for a JS-DFT trial WF.
The results, as obtained employing the modified TABC method in conjunction with a JS-DFT trial WF, are reported in Fig.~\ref{fig:TABC-E_ctf}.
%In the atm-bcc phase, the selected cutoff energies were not sufficient to reach the convergence, as one could expect since the metallic systems are usually more demanding in terms of $E_{\text{ctf}}$.
%In contrast, $10~\text{Ry}$ is an adequate cutoff energy for the mol-hcp phase.
As can be seen, the convergence with respect to $E_{ctf}$ is much slower for the metallic atm-bcc than for the insulating mol-hcp phase of solid hydrogen, where as few as $10~\text{Ry}$ is adequate. Moreover, in all cases $n_K=5$ is sufficient to consider all finite size effects for $N$ larger than $16$. Nevertheless, since the accumulated statistics for each $K$-point contribute to the overall average, the total computational cost is essentially independent from $n_K$.  
%requires more space on the hard disk, but it does not demand longer computational times nor larger RAM within VMC.
\begin{figure}%[htbp]
	\centering
		\includegraphics[width=1.0\columnwidth]{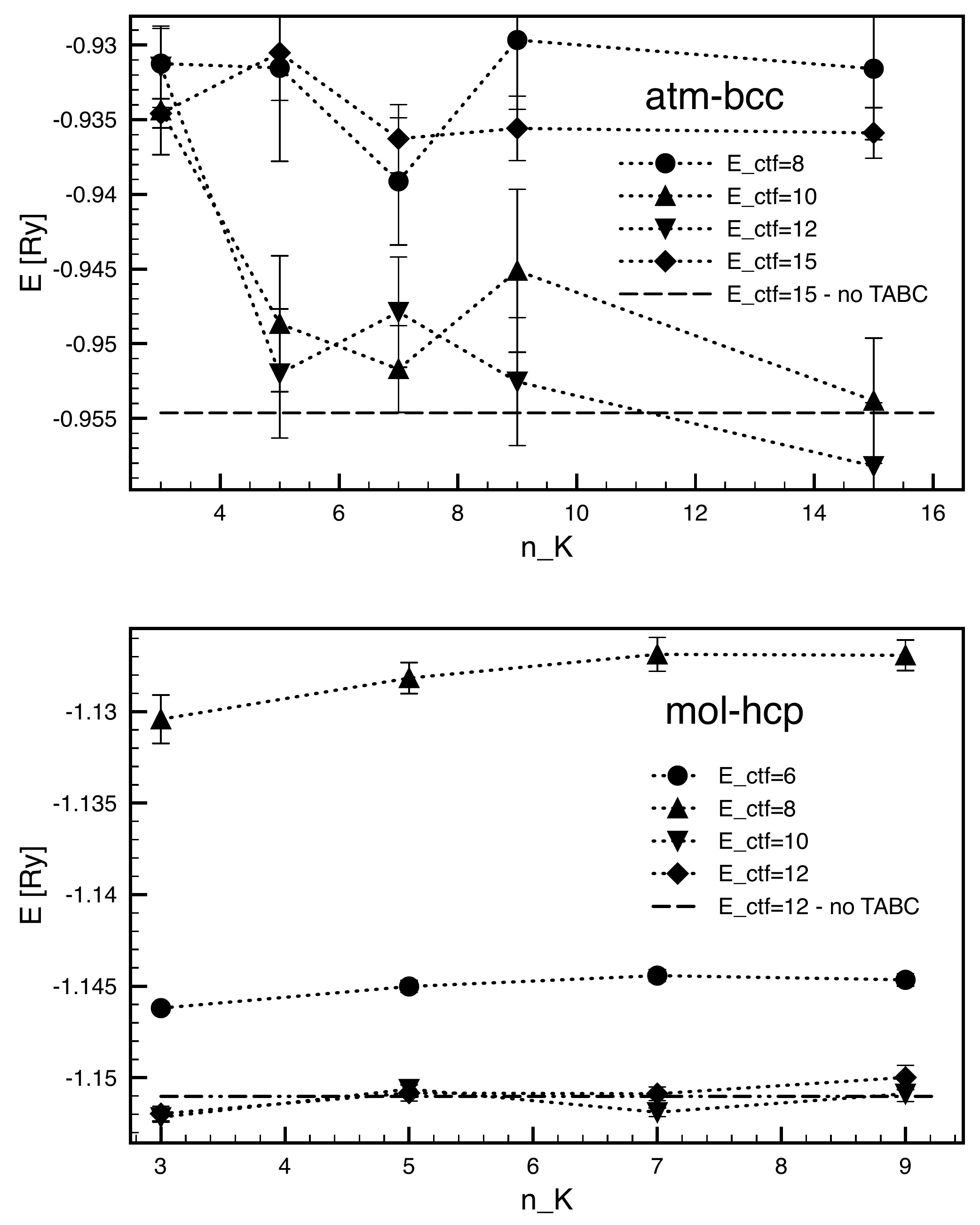}
	\caption{Variational energy of solid hydrogen in the atm-bcc phase at $r_s=1.31$ and mol-hcp phase at $r_s=2.61$ as a function of $E_{\text{ctf}}$ and $n_K$. The energies were calculated for $N=16$ using the JS-DFT trial WF.}
	\label{fig:TABC-E_ctf}
\end{figure}

The effectiveness of the modified TABC approach as a function of $N$ is demonstrated in Fig.~\ref{fig:convergence_N}. As can be seen, the TABC provide a quicker convergence to the thermodynamic limit especially in the case of the metallic atm-bcc phase that obeys rather large finite size effects.
\begin{figure}%[htbp]
	\centering
		\includegraphics[width=1.0\columnwidth]{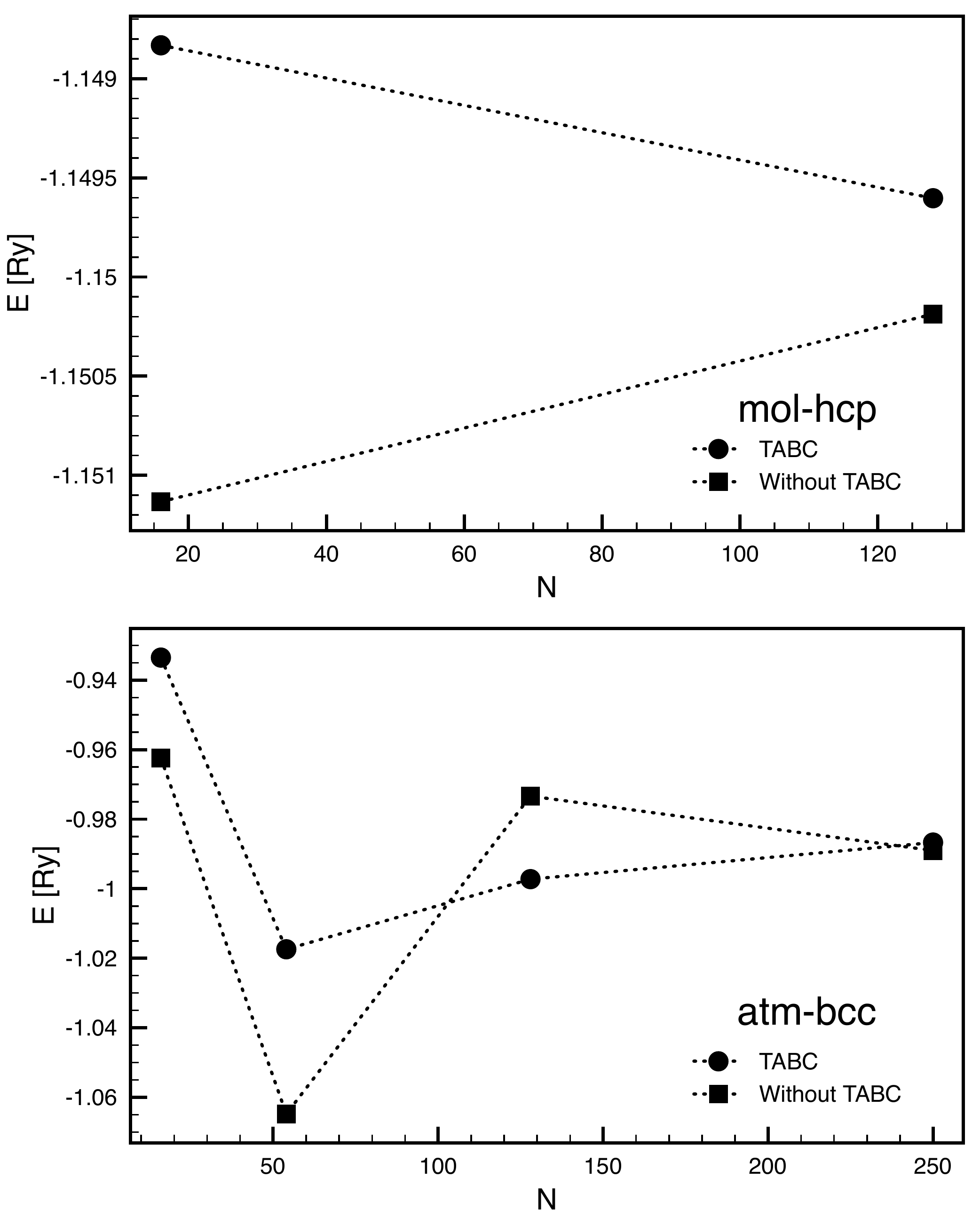}
	\caption{Variational energy for solid hydrogen in the atm-bcc phase at $r_s=1.31$ and mol-hcp phase at $r_s=2.61$ as a function of $N$. The energies were calculated for $N=16$ using the JS-DFT trial WF.}
	\label{fig:convergence_N}
\end{figure}

\section{Results and Discussion} \label{sec:results}

To demonstrate the predictive power of the SWF in general and the ASWF-DFT trial WF in particular, we investigate the metal-insulator-transition (MIT) from the metallic atm-bcc to the insulating mol-hcp phase of solid hydrogen. The corresponding results using the conventional JS trial WF are shown in Fig.~\ref{BCCvsHCP-assessment}. Not surprisingly, using the JS-DFT trial WF, the variational energies are throughout more favorable than the ones obtained by the JS-pw trial WF. However, while the latter are in reasonable good agreement with the former for the metallic atm-bcc phase, the JS-pw trial WF fails to describe the insulating mol-hcp phase. In general, the results of the JS-pw and JS-DFT trial WFs are deviating from each other with increasing distance between the monomers that implies with larger multireference character. Interestingly, we find that especially for large monomer separation the rather simple JS-bi-atomic and JS-1s {trial} WFs are in fact even more accurate than the JS-DFT results. Considering its simplicity, the JS-1s {trial} WF performs relatively well for both of the considered phases. 
\begin{figure}
\centering
\includegraphics[width=1.0\columnwidth]{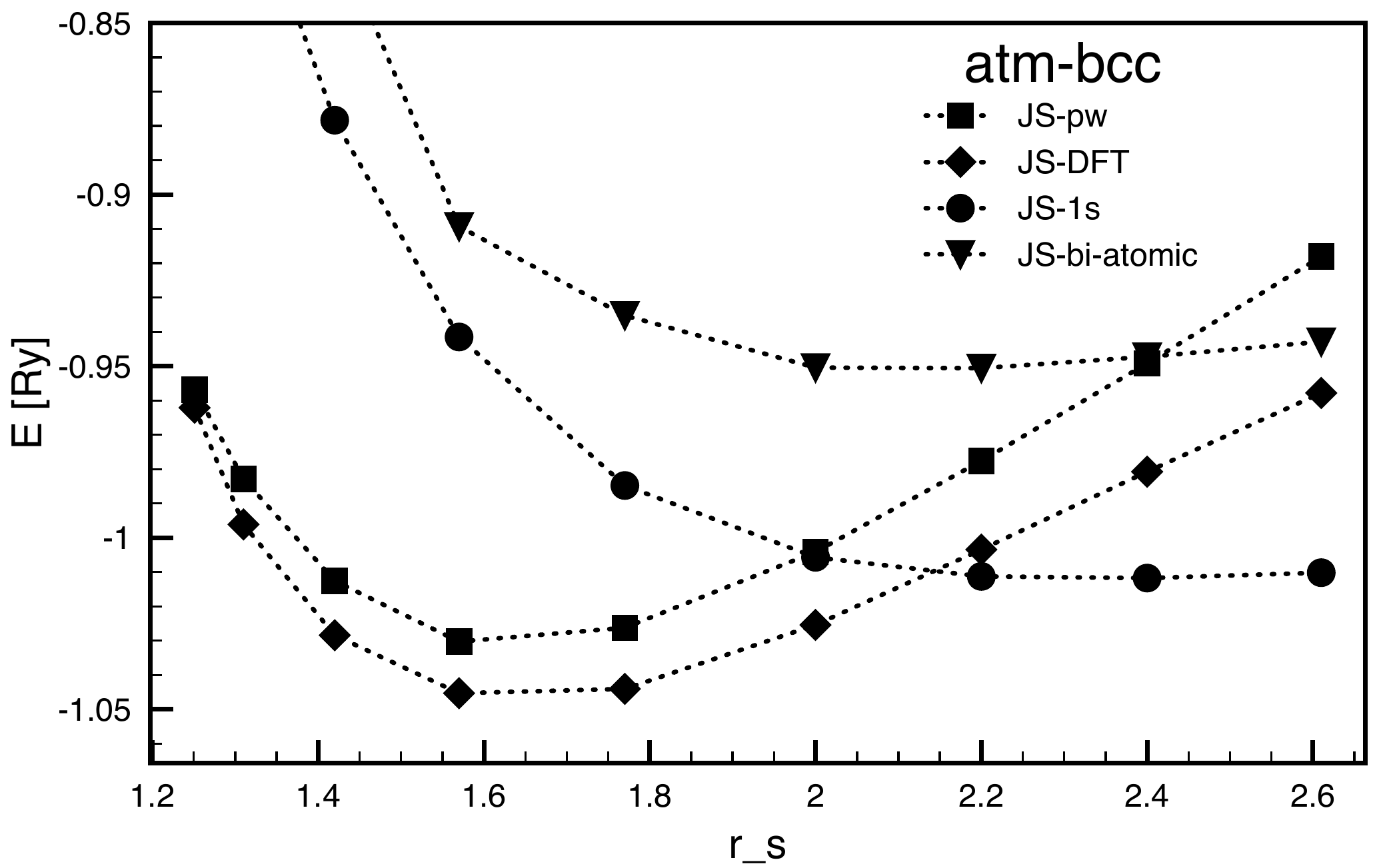}
\includegraphics[width=1.0\columnwidth]{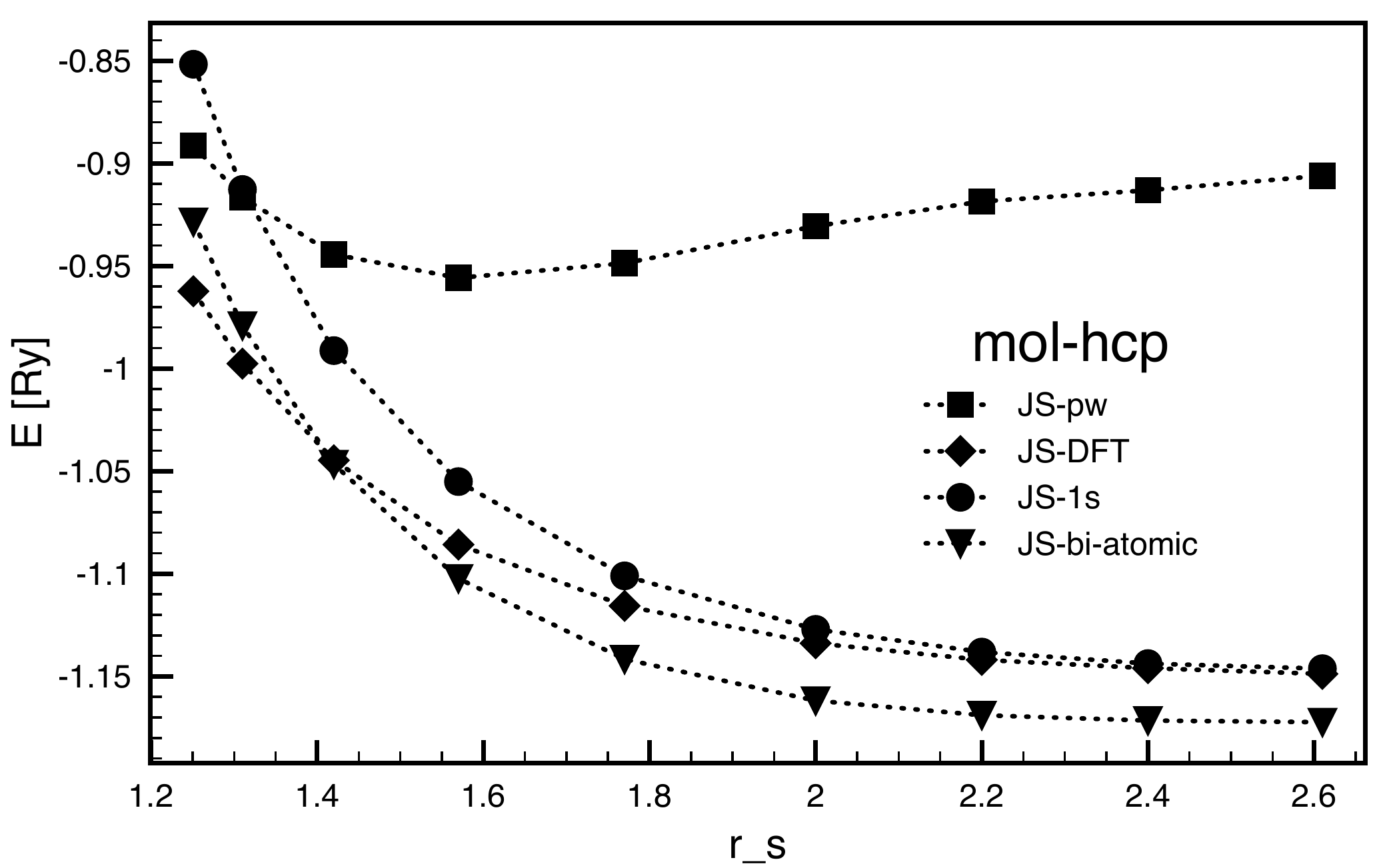}
\caption{
\label{BCCvsHCP-assessment}
Variational energies of the metallic atm-bcc and the insulating mol-hcp phases of solid hydrogen using various JS-type trial WFs.} 
\end{figure}
However, as can be seen in Fig.~\ref{BCCvsHCP}, the increased accuracy of the ASWF with respect to the JS-type WFs is rather limited. Although, the improvement is noticeable in the case of the JS-pw trial WF, for the more accurate JS-DFT approach it renders inessential. This is to say that the observed improvement in the employed WF is nearly entirely due to the application of DFT to construct the SD, which subsequently is not further enhanced by the present shadow formalism. The latter suggests that the eventual DFT-based trial WFs are already very accurate. 
\begin{figure}
\centering
\includegraphics[width=1.0\columnwidth]{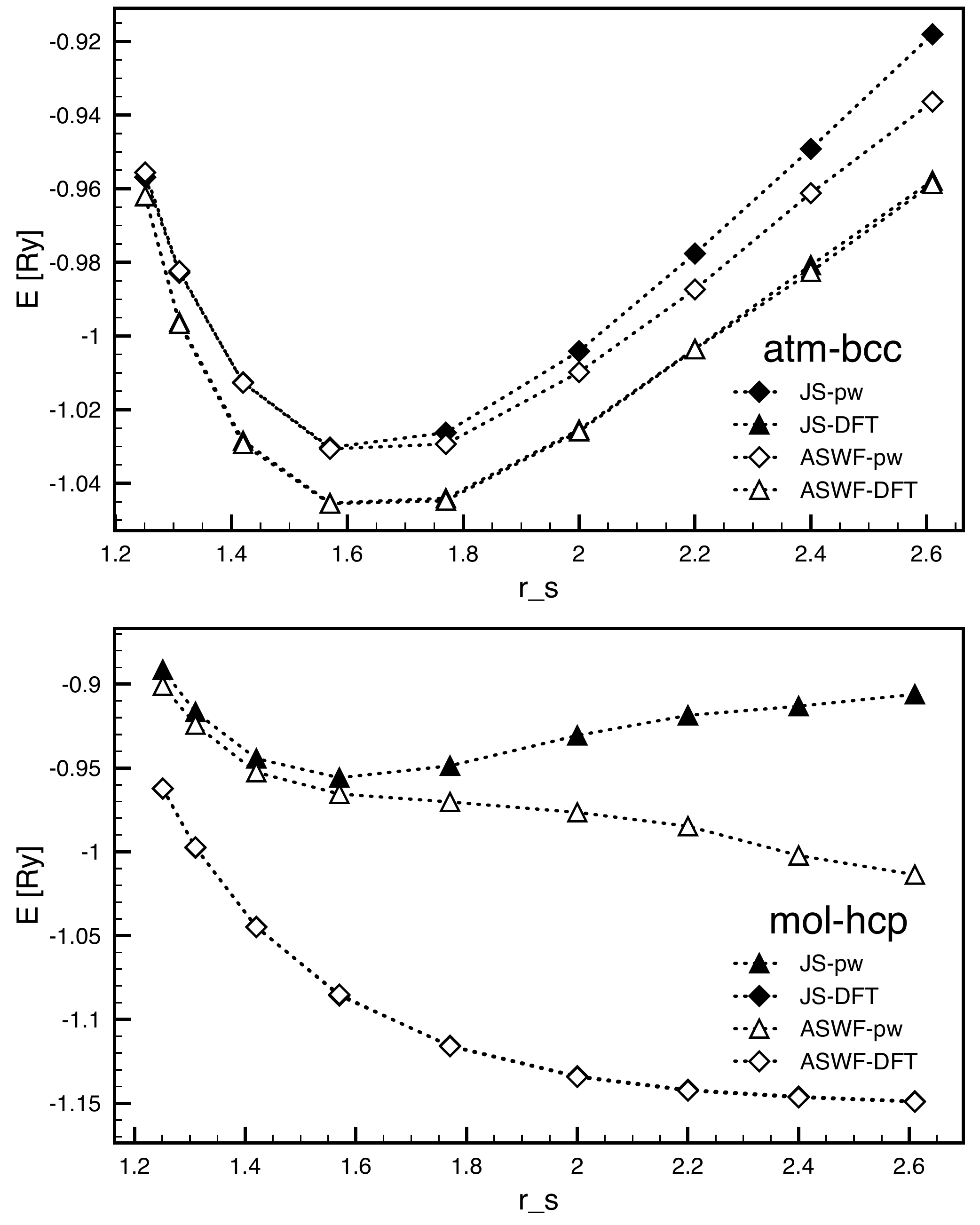}
\caption{
\label{BCCvsHCP}
Variational energies of the metallic atm-bcc and the insulating mol-hcp phases of solid hydrogen using the ASWF and JS-type trial WFs.} 
\end{figure}

In order to the determine the transition pressure of the MIT for the various {trial} WF investigated here, in Fig.~\ref{MIT} the energies the metallic atm-bcc and the insulating mol-hcp phases are shown as a function of $r_s$. 
\begin{figure}
\centering
\includegraphics[width=1.0\columnwidth]{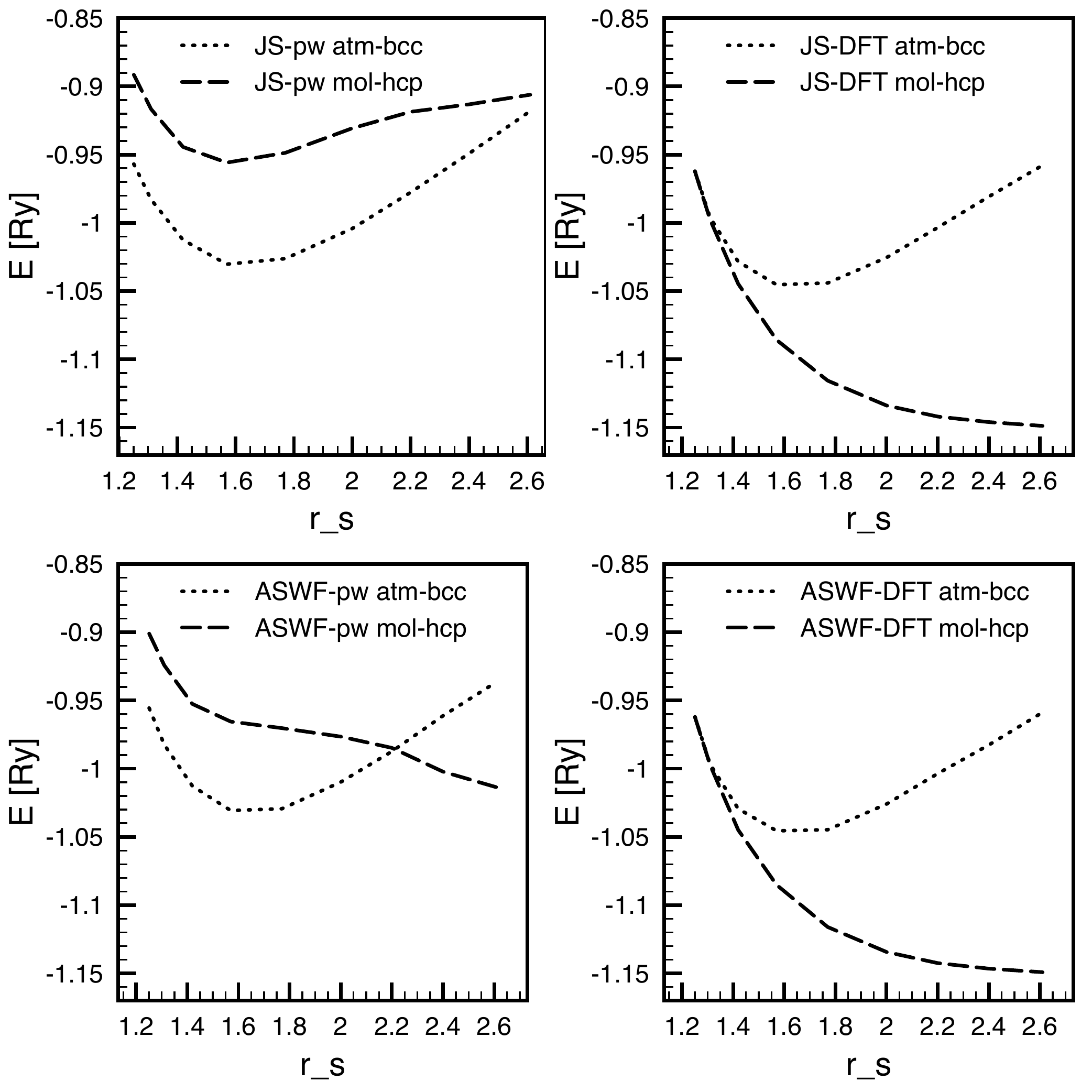}
\caption{
\label{MIT}
The MIT between the metallic atm-bcc and the insulating mol-hcp phases of solid hydrogen using the ASWF and JS-type trial WFs.} 
\end{figure}
Using the common tangent construction, we find an MIT pressure of 12~GPa for the JS-pw trial WF, which is even lower than predicted by Wigner and Huntington back in 1935 \cite{JCP.3.764, Nellis2013}. Applying the more accurate ASWF formalism instead of the plain JS trial WF, the MIT pressure slightly increases to and 45~GPa. 
%The energies of the bcc and hcp get very close at $r_s=1.31$ and become equal within the error bars at $r_s=1.251$. One can estimate the pressure of hydrogen using the expression
%\begin{equation}
%   P = - \frac{\partial U}{\partial V} = - \left( \frac{\partial v}{\partial r_s} \right)^{-1} \left( \frac{\partial u}{\partial r_s} \right)
%\end{equation}
%where $U$ is the internal energy, $u$ the internal energy per atom, $V$ the volume of the system, and $v$ the volume per atom. 
%Then, using the energies of the JS-DFT trial WF,
%\begin{equation}
%   \frac{\partial u}{\partial r_s} = \frac{E(r_s=1.31) - E(r_s=1.251)}{1.31 - 1.251} =  -0.5770 \pm 0.0005 \; \text{Ry}  %\frac{-0.99613524 + 0.9620947}{0.059}
%\end{equation}
%and
%\begin{equation}
%   \left( \frac{\partial v}{\partial r_s} \right) = 4 \pi r_s^2 %=15.7205296386
%\end{equation}
%As a result, we can estimate the transition pressure at $r_s=1.251$ as
%\begin{equation}
%   P \simeq 540 \text{GPa} %=0.03670359798 Ry/Bohr^3
%\end{equation}
However, as before, substituting the pw orbitals within the SD by those of a mean-field DFT calculation, results not only in a substantially reduced variational energy, but also in a dramatically increased MIT pressure. Specifically, employing the JS-DFT trial WF results in a transition pressure of 395~GPa, while the usage of the present ASWF transformation increases the MIT pressure to even 520~GPa, which is still beyond the largest pressures experimentally realized so far at low temperature. Therefore, although the variational energy is only slightly improved by the ASWF when using DFT orbitals in the SD, the impact on the transition pressure is rather large. Moreover, the present results immediately suggest the general trend that more accurate the employed trial WF, the higher the resulting MIT pressure. In fact, despite the simplicity of the underlying JS-type trial WF, the present ASWF-DFT results compares relatively favorable with recent state-of-the-art finite-temperature QMC calculations using much more sophisticated {trial} WF \cite{Morales2010, Liberatore2011, Mazzola2014, PhysRevLett.114.105701, Pierleoni2016}. Nevertheless, it is important to note that the here considered solid phases of insulating molecular and metallic atomic hydrogen are not the energetically most favorable structures known to date and as such only qualitative representatives of the MIT \cite{RevModPhys.84.1607}. Furthermore, the possible existence of a quantum fluid phase at zero temperature}, 
{which is consistent with a maximum in the melting curve} \cite{PhysRevB.61.6535, 2003PNAS..100.3051S, 2004Natur.431..669B, PhysRevLett.100.155701, 2009JETPL..89..174E}, is neglected. 

\section{Conclusion}

In conclusion, we have extended the ASWF to periodic large-scale systems made up fermions. For that purpose, we have exploited an improved SR scheme to efficiently optimize the employed ASWF \cite{CalcavecchiaEPL2015}, and combined it with enhanced PC and TABC techniques.
To demonstrate the predictive power of this approach, we investigated the MIT of solid hydrogen at very high pressure. In particular we found that the ameliorated accuracy of the ASWF results in a significantly increased transition pressure {of 520~GPa}. 
%Speaking about ZPE, the remaining factor is its influence on the relative stability of different structures. Although in absolute value the ZPE is rather large, contrary to solid atomic hydrogen, there is a strong tendency in the molecular case that the ZPE of different structures cancel each other out, which is even more pronounced with increasing pressure \cite{PhysRevLett.74.1601}. In any case, the fact that the ZPE tends to favor symmetric structures \cite{PhysRevB.36.2092, PhysRevLett.38.415} further strengthens our prediction in favor of the P6$_{3}$/m phase as the most likely candidate for Phase III. 

\begin{acknowledgments}
The authors would like to thank the Graduate School of Excellence MAINZ for financial support and Markus Holzmann for useful comments. The Gauss Center for Supercomputing (GCS) is kindly acknowledged for providing computing time through the John von Neumann Institute for Computing (NIC) on the GCS share of the supercomputer JUQUEEN at the J\"ulich Supercomputing Centre (JSC).
\end{acknowledgments}

\bibliography{tdk} % References file

\end{document}